\documentclass[a4paper,11pt]{article}
\pdfoutput=1 

\usepackage{jheppub} 

\usepackage{color}
\usepackage{graphicx}
\usepackage{amsmath}
\usepackage{amssymb}
\usepackage{xspace}
\usepackage[small]{subfigure}

\newlength{\fighskip} \fighskip=2pt
\newlength{\figvskip} \figvskip=3pt

\newcommand*{\figbox}[2]{{
  \def\figscale{#1}
  \def\arraystretch{0.8}
  \arraycolsep=0pt
  \begin{array}{c}
    \vbox{\vskip\figscale\figvskip
      \hbox{\hskip\figscale\fighskip
        \includegraphics[scale=\figscale]{#2}}}
  \end{array}}}

\title{Hayden-Preskill protocol and decoding Hawking radiation at finite temperature}
\author[a,b]{Ran Li,}
\author[b,c,*]{Jin Wang \note[*]{Corresponding author}}
\affiliation[a]{School of Physics, Henan Normal University, Xinxiang 453007, China}
\affiliation[b]{Department of Chemistry, Stony Brook University, Stony Brook, NY 11794, USA}
\affiliation[c]{Department of Physics and Astronomy, Stony Brook University, Stony Brook, NY 11794, USA,}

\emailAdd{liran@htu.edu.cn}
\emailAdd{jin.wang.1@stonybrook.edu}

\abstract{We study the Hayden-Preskill thought experiment at finite temperature and obtain the decoupling condition that the information thrown into an old black hole can be extracted by decoding the Hawking radiation. We then consider the decoding Hayden-Preskill protocol at finite temperature assuming the observer outside the black hole who has the access to the full radiation and the unitary dynamics of the black hole. We also consider the cases when the Hawking radiation has noise and decoherence in the storage. The decoding probabilities and the corresponding fidelities are calculated. It is shown that for all the three cases we have considered, the decoding fidelities are less than unity in general. This result indicates that at finite temperature, the decoding strategy and the recovery algorithm is harder to realize than that at infinite temperature.}

\begin{document}

\maketitle

\section{Introduction} 

Hawking's semiclassical calculation \cite{Hawking:1975vcx} indicates that the black holes radiate thermally and therefore should evaporate completely. The consequence is that the initial black hole, which can be taken as a pure state, eventually evolves to a mixed state \cite{Hawking:1976ra}. This violates the fundamental principles of quantum mechanics, and gives rise to the black hole information paradox. One can refer to \cite{Harlow:2014yka} for a recent review on this aspect.

If one assumes that the dynamics process of the black hole evolution is unitary, the information of the matter that collapses to the black holes or the information that swallowed by the black holes should be leaked out from it in the form of Hawking radiation. However, this could lead to the cloning of quantum state, which is contradiction to the linearity of quantum mechanics \cite{Preskill:1992tc}. The black hole complementary principle \cite{Susskind:1993if} proposed by Susskind et al. argued that the observers inside the event horizon and outside cannot compare their quantum states and this does not violate the quantum non-cloning theorem. The argument is based on the Page's work of the entanglement entropy of Hawking radiation from the evaporating black holes \cite{Page:1993df,Page:1993wv}, which reveals that the information would be released from the black hole after the Page time when the black hole entropy is equal to the radiation entropy. The Hayden-Preskill thought experiment \cite{Hayden:2007cs} addressed the question that the information swallowed by the old black hole can be extracted very quickly if assuming the black hole as being the fastest scrambling system. The Hayden-Preskill thought experiment has stimulated much works on the black hole information paradox, including the AMPS thought experiment \cite{Almheiri:2012rt}, the fast scrambling dynamics in quantum many body system \cite{Sekino:2008he,Lashkari:2011yi,Shenker:2013pqa,Maldacena:2015waa}, the traversable wormholes \cite{Gao:2016bin,Maldacena:2017axo} and so on.

Recently, Yoshida and Kitaev \cite{Yoshida:2017non} considered the strategy of decoding and reconstructing the quantum state from the Hawking radiation in the Hayden-Preskill thought experiment. The decoding strategy indicates that, if the dynamics of the black hole is unitary, one should be able to recovery the quantum state that is thrown into black hole from the Hawking radiation that comes out. Some related works can be found in \cite{Yoshida:2018vly,Yoshida:2018ybz,Yoshida:2019qqw,Cheng:2019yib,Yoshida:2019kyp,Bao:2020zdo,Schuster:2021uvg,Hayata:2021kcp,Yoshida:2021xyb}. In particular, the decoherence and noise effects on the random unitary dynamics or on the storage of Hawking radiation are considered in \cite{Yoshida:2018vly} and \cite{Bao:2020zdo}. The decoding strategy of Hawking radiation in the Hayden-Preskill thought experiment has also been generalized to study the interior operator of black hole \cite{Yoshida:2018ybz,Yoshida:2019kyp}, the implications in the black hole firewall paradox \cite{Yoshida:2019qqw}, finite temperature case \cite{Cheng:2019yib,Yoshida:2019kyp}, and the Clifford unitary dynamics case \cite{Yoshida:2021xyb}. Besides the strong interest in the black hole information paradox, such progress may stimulate the researches on quantum chaos and quantum information scrambling \cite{Hosur:2015ylk} in many body quantum systems.

In this paper, we will study the more realistic case by considering the black hole as being thermal. Generally speaking, according to the calculation of Hawking, a black hole should be treated as a thermal system at finite temperature. Therefore, in the Hayden-Preskill thought experiment, the black hole should be maximally entangled with the Hawking radiation in the thermal field double (TFD) state at fixed finite temperature after the Page time \cite{Wald:1975kc}. However, this is also a simplified assumption because the temperature of the black hole varies with time during the evaporating process. But this assumption can also be reasonable due to the well known example that the the temperature of CGHS black hole \cite{Callan:1992rs} is a constant which does not depend on the black hole mass. Under this assumption, We will reconsider the decoupling condition \cite{Hayden:2008os} that the information of the message system can be extracted from the Hawking radiation. We will also consider the recovery strategy of the quantum information from the Hawking radiation without/with noise and decoherence in the storage.

The rest of the paper is arranged as follows. In Section \ref{HP_finite_tem}, we consider the Hayden-preskill thought experiment at finite temperature and obtain the decoupling condition for the recoverability of quantum information from decoding the Hawking radiation. In Section \ref{Decoding_HR}, we consider the decoding strategy of Hawking radiation at finite temperature and calculate the probability and the fidelity of the decoding process. In Section \ref{decoding_HR_ERR} and Section \ref{Decoding_HR_dep}, we consider the decoding strategy when the erased errors and the quantum depolarizing channel in the storage of the Hawking radiation are taken into account, and calculate the corresponding decoding probabilities and the fidelities. We conclude this paper with the discussions in Section \ref{conclusion}.      

\section{Hayden-Preskill thought experiment at finite temperature}\label{HP_finite_tem} 

In this section, we consider the Hayden-Preskill thought experiment at finite temperature. In particular, we introduce the graphical representation developed by Yoshida and Kitaev \cite{Yoshida:2017non} and use it to derive the decoupling condition at the finite temperature by assuming that the dynamics is governed by the random unitary operator. 

Firstly, let us introduce the thought experiment of Hayden and Preskill \cite{Hayden:2007cs}. In the Hayden-Preskill thought experiment, one starts with the system which is composed by the black hole $B$ and its radiation $B'$. The black hole $B$ is maximally entangled with its radiation $B'$, which implies that the black hole have already emitted half of its qubits, i.e. the black hole has already pasted its Page time. Suppose that Alice has a message which is represented by the system $A$, and the message system $A$ is also maximally entangled with the reference system $R$. At this time, the message system $A$ is thrown into the black hole. See the graphical representation of the Hayden-Preskill thought experiment in Eq.(\ref{psi_HP}). Then, the black hole $B$ is maximally entangled with the system $R$ and $B'$. The whole process of absorbing the message system $A$ and the consequent radiating can be modeled by the random unitary operator $U$. After some time, the black hole $B$ will evolve to the late radiation system $D$ and the remnant black hole $C$. There is another observer Bob, who has access to the early radiation $B'$ and the late radiation $D$. Once $D$ is large enough, the correlations between the black hole $B$ and the reference system $R$ disappear. Since the overall state is pure, the reference system $R$ is purified by the radiation system $B'$ and $D$. In principle, the information of the message system $A$ can be extracted from decoding the Hawking radiation. It is shown that the observer Bob is able to recover the information by only collecting a few more radiated qubits than the qubits of the message system $A$. This results indicates that the old black holes, which have already proceeded past the half way point of the evaporation, are the mirrors that release quantum information very rapidly in the form of Hawking radiation. This work has led to the studies of the fast scrambling conjecture of the black holes which claims that the black holes are the fastest information scramblers \cite{Sekino:2008he,Lashkari:2011yi,Shenker:2013pqa,Maldacena:2015waa}.

The previous discussions are based on the assumption that the old black hole is maximally entangled with the early radiation in the  Einstein-Podolsky-Rosen (EPR) state. Here, we consider the case where the black hole $B$ and its early radiation $B'$ are in a thermofield double (TFD) state at finite temperature. In fact, the EPR state can be considered as the infinite temperature limit of the TFD state. Our motivation is that the black hole is generally a thermal system at finite temperature. In general, the Hawking temperature of the black hole during the evaporation process will be a time dependent quantity. However, it is known that the CGHS black hole \cite{Callan:1992rs} is a special example where the temperature is a constant and does not depend on the black hole mass. It is also true for the two dimensional dilaton black holes in RST model \cite{Russo:1992ax}, in BPP model \cite{Bose:1995pz}, and in a generalized model \cite{Cruz:1997nj,Cruz:1995zt,Cruz:1996pg}. Therefore, without the loss of generality, it is reasonable to take the initial state of the black hole $B$ and its radiation $B'$ to be a TFD state at finite temperature in the Hayden-Preskill thought experiment. We consider the case that the message systems $A$ and the reference system $R$ are in the EPR state. We will discuss the decoupling condition that the information in the message system can be extracted by decoding the Hawking radiation.

Now we consider the Hayden-Preskill thought experiment at finite temperature. Let $d_A$, $d_B$, $d_C$, $d_D$, $d_R$ and $d_{B'}$ be the Hilbert space dimensions of the message system $A$, the initial black hole system $B$, the remnant black hole $C$, the late radiation $D$, the reference system $R$, and the early radiation $B'$, respectively. The dimension of the total Hilbert space is $d^2$ with $d=d_Ad_B=d_Cd_D=d_Rd_{B'}$. With the prepared states for the system $BB'$ and $RA$, after the time evolution, the state of the system is given by
\begin{eqnarray}\label{psi_HP}
|\Psi_{HP}\rangle=(I_{R}\otimes U_{AB}\otimes I_{B'}) |EPR\rangle_{RA}\otimes |TFD\rangle_{BB'}=\figbox{0.3}{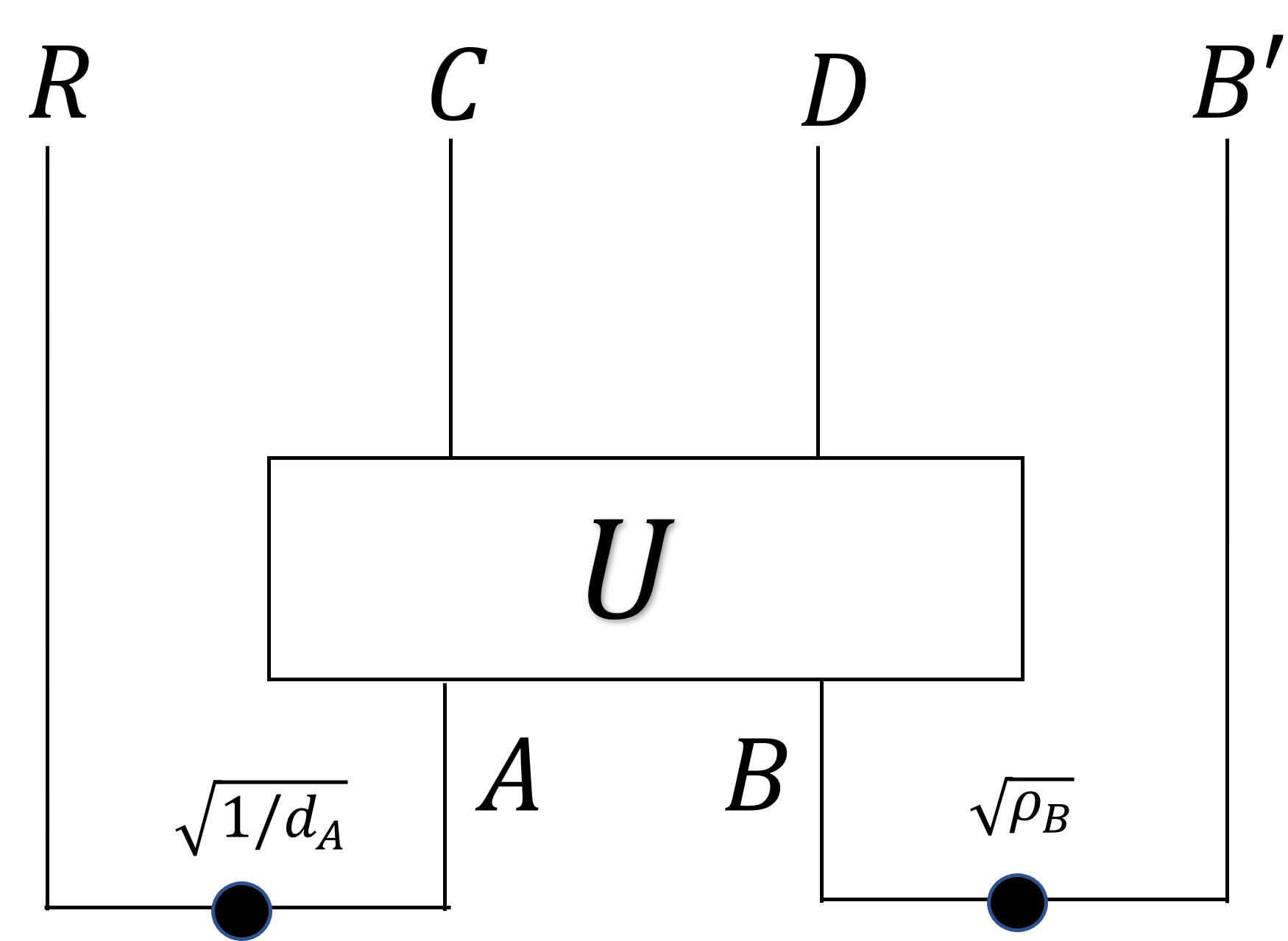} 
\end{eqnarray}
where the whole process of the message system $A$ swallowed by the black hole and the consequent radiating is modeled by the random unitary operator $U$. Because the whole process is a complex one, it is plausible to assume that the system $CD$ is random. In the above equation, the EPR state of the system $AA'$ and the TFD state of the system $BB'$ are defined as 
\begin{eqnarray}
|EPR\rangle_{RA}&=&\frac{1}{\sqrt{d_A}}\sum_{n} |i_{R}\rangle \otimes |i_{A}\rangle = \figbox{0.4}{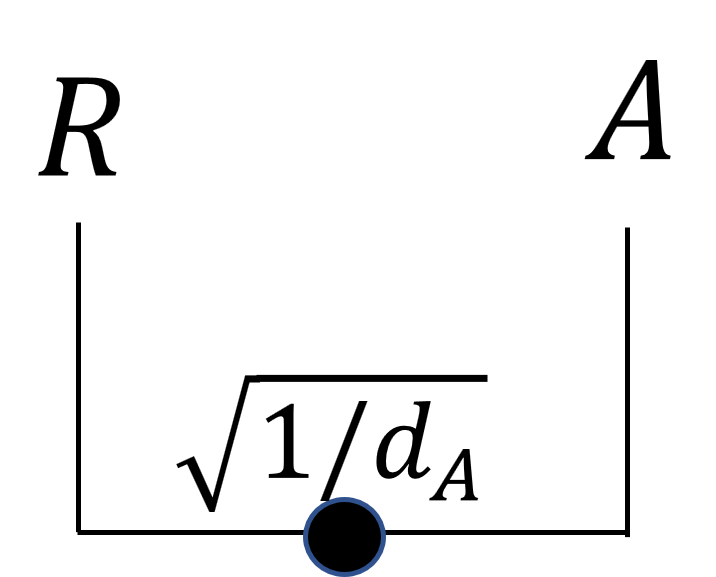}\;\\
|TFD\rangle_{BB'}&=&\frac{1}{\sqrt{Z_B}}\sum_{n}e^{-\beta_B E^B_n/2} |n_{B}\rangle \otimes |n_{B'}\rangle = \figbox{0.4}{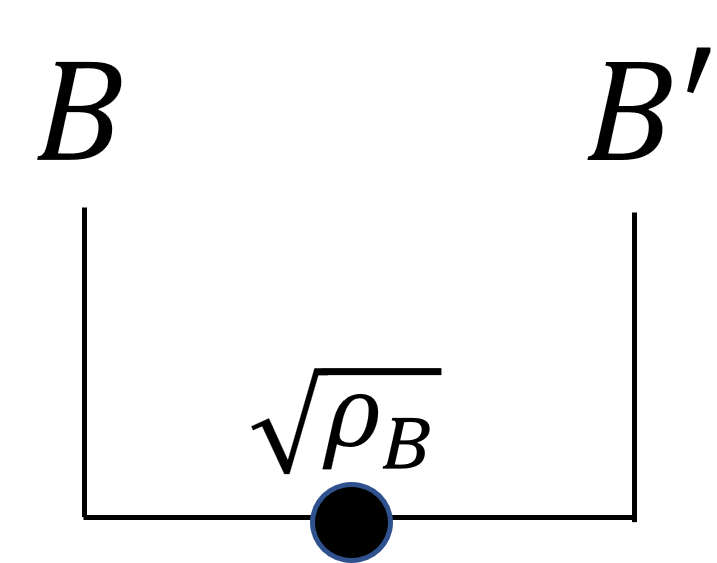}\;. 
\end{eqnarray}
Note that $\beta_B$ is the inverse temperature of the system $B$ and $B'$, $Z_B=\sum_{n}e^{-\beta_B E^B_n}$ is the normalization coefficient, $|n\rangle$ is the eigenstate of the system $B$ with energy $E^B_n$. The reduced density matrix for the system $B$ is just the thermal density matrix $\rho_B=e^{-\beta_B H_B}/Z_B$. In the above graphical representation, for the EPR state the black dot represents the normalization factor $\frac{1}{\sqrt{d_A}}$. For the TFD state, the black dot represents the factor $\sqrt{\rho_B}=\frac{1}{\sqrt{Z_B}} e^{-\beta_B H_B/2}$. The normalization condition of the TFD state can be represented as
\begin{eqnarray}
\langle TFD |TFD\rangle_{BB'} = \figbox{0.4}{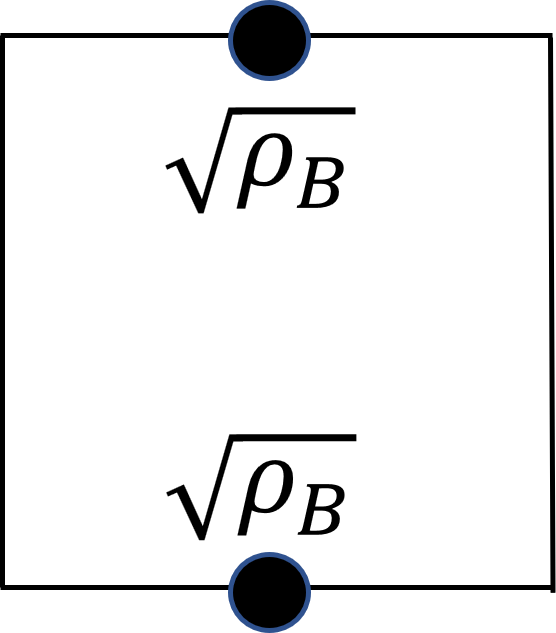} = \figbox{0.4}{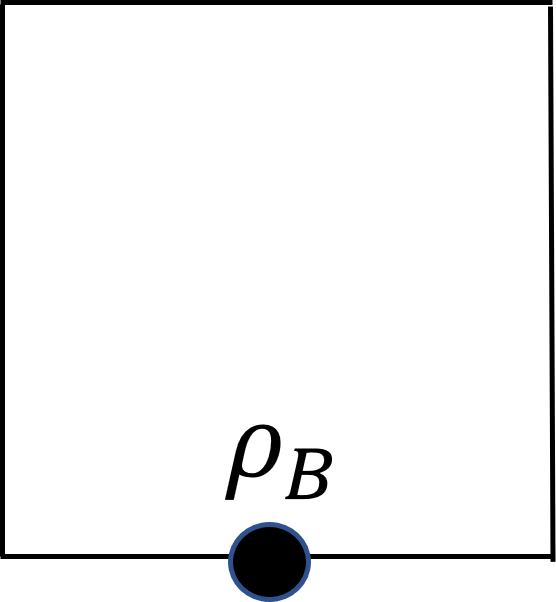}=\textrm{Tr} \rho_B=1\;. \end{eqnarray}

It should be noted that there is a certain technical assumption which seems to be problematic for physical reason \cite{Yoshida:2017non}. For the given input ensemble $\rho_{AB}=\rho_A\otimes \rho_B$, it is assumed that the output ensemble factorizes into $C$ and $D$, i.e. $\rho_{CD}=U\rho_{AB}U^\dagger\approx \rho_C\otimes \rho_D$. Because the black hole evaporates adiabatically, this assumption is equivalent to that there is no interaction between the two subsystems $C$ and $D$. Although, the Hawking radiation eventually decouples from the black hole, it is impossible to draw a sharp line between the two subsystems for factoring the the density matrix $\rho_{CD}$. However, as discussed in the Appendix of \cite{Yoshida:2019kyp}, it is possible to discard the short range entanglement between $C$ and $D$ and construct the subspace $C_{sub}$ and $D_{sub}$ on which the density matrix of $CD$ can be factorized into $ \rho_{C_{sub}}\otimes \rho_{D_{sub}}$. Therefore, we will proceed with this assumption to discuss the finite temperature case.

The corresponding density matrix for the state $|\Psi_{HP}\rangle$ can be graphically represented by 
\begin{eqnarray}
|\Psi_{HP}\rangle \langle\Psi_{HP}|=\figbox{0.3}{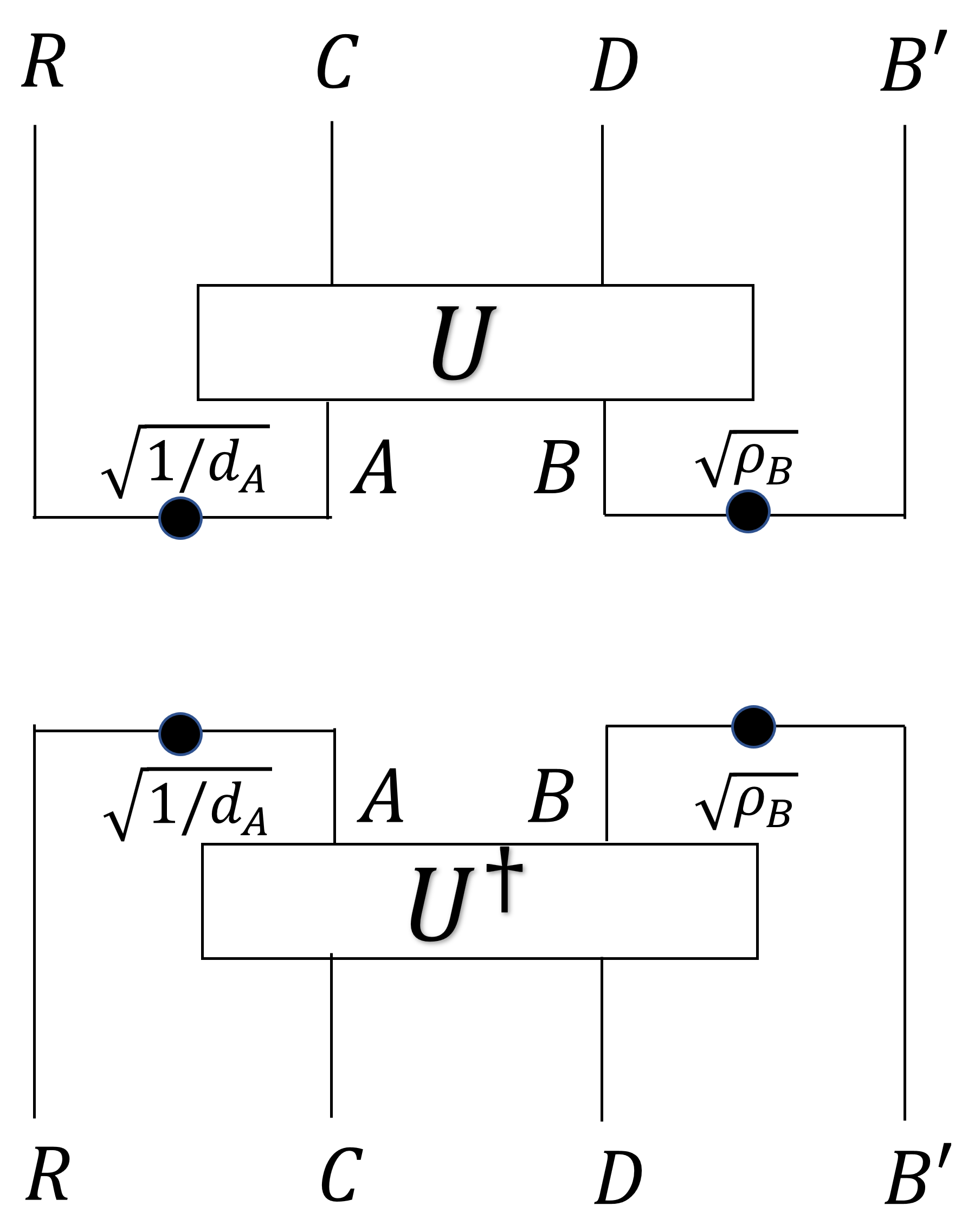} 
\end{eqnarray}

The reduced density matrix for the subsystem $RC$ is then given by
\begin{eqnarray}\label{reduced_RC}
\rho_{RC}=\textrm{Tr}_{D,B'}|\Psi_{HP}\rangle \langle\Psi_{HP}|=\figbox{0.3}{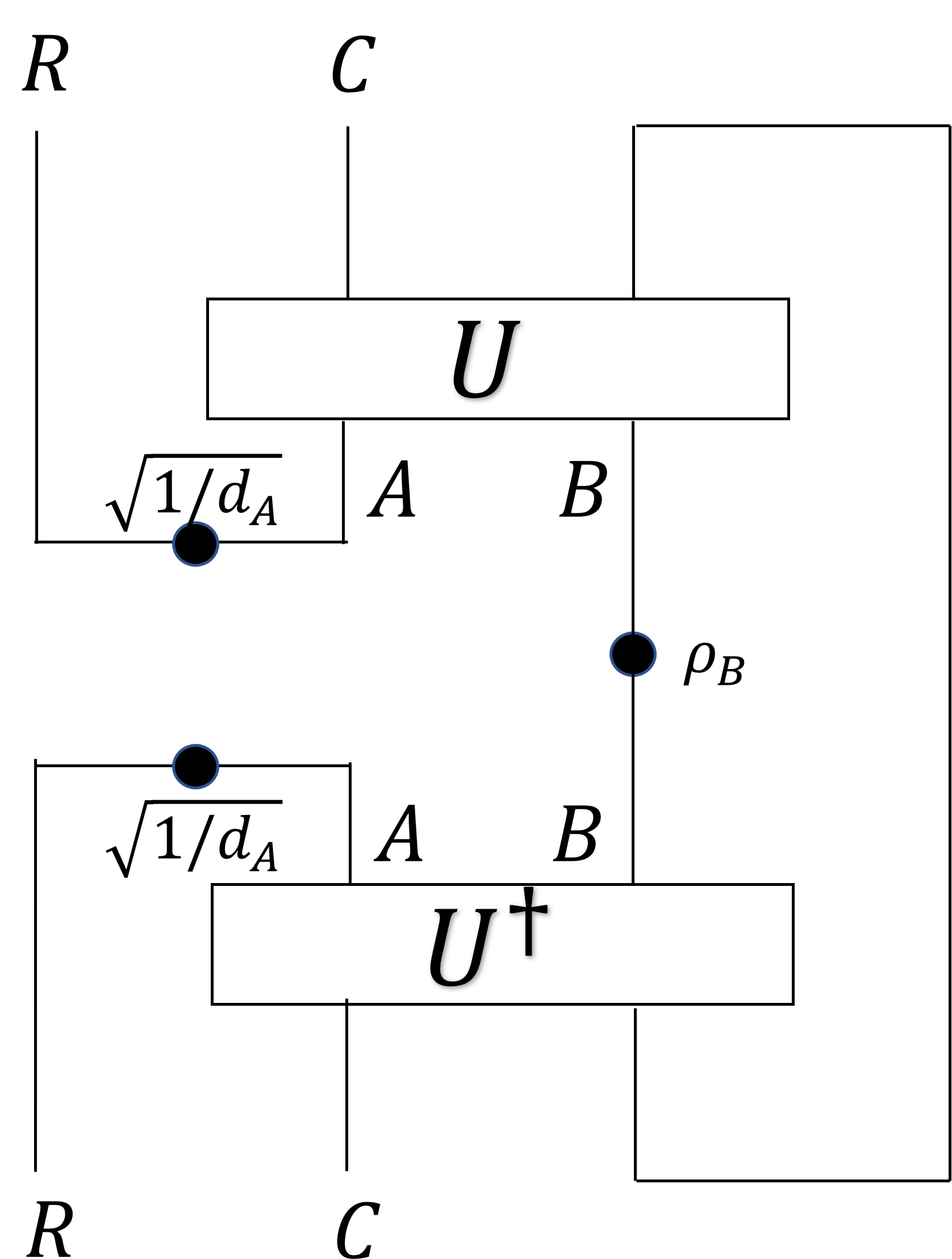} 
\end{eqnarray}
It can be easily checked that the reduced density matrix of the system $RC$ is normalized, i.e. $\textrm{Tr} \rho_{RC}=1$.

Our goal is to discuss the condition under which the reference system $R$ and the remnant black hole $C$ are decoupled. This can be achieved by estimating the operator distance \cite{Nielsen:2000} between the reduced density matrix $\rho_{RC}$ and the product state of the system $R$ and $C$ averaged over the random unitary matrix $U$. We should consider the following quantity 
\begin{eqnarray}\label{quantity}
\left(\int dU \|\rho_{RC}-\frac{1}{d_Rd_C} I_R \otimes I_C \|_1\right)^2\;,
\end{eqnarray}
where $\frac{1}{d_R} I_R$ and $\frac{1}{d_C} I_C$ are the maximally mixed density matrices of the system $R$ and $C$. The operator trace norm $\|\cdot\|_1$, which is also called the $L_1$ norm, is defined for any operator $M$ as $\|M\|_1=\textrm{Tr}\sqrt{M^\dagger M}$. If the quantity in Eq.(\ref{quantity}) is small enough, the correlations between the reference system $R$ and the remnant black hole $C$ can be ignored. Therefore, we should try to estimate the upper bound of the quantity in Eq.(\ref{quantity}).

By defining the $L_2$ norm as $\|M\|_2=\sqrt{\textrm{Tr} M^\dagger M}$, and using the inequality $\|M\|_2\leq\|M\|_1\leq \sqrt{N} \|M\|_2$ with $N$ being the dimensionality of the Hilbert space, one can estimate
\begin{eqnarray}
\left(\int dU \|\rho_{RC}-\frac{1}{d_Rd_C} I_R \otimes I_C \|_1\right)^2
&\leq& \int dU \|\rho_{RC}-\frac{1}{d_Rd_C} I_R \otimes I_C \|_1^2\nonumber\\
&\leq& d_Rd_C \int dU \|\rho_{RC}-\frac{1}{d_Rd_C} I_R \otimes I_C \|_2^2\nonumber\\
&=&d_Rd_C \int dU \textrm{Tr}(\rho_{RC})^2-1\;,
\end{eqnarray}
where we have used Jensen's inequality and the fact that $\textrm{Tr} \rho_{RC}=1$.

Using Eq.(\ref{reduced_RC}),  $\textrm{Tr}(\rho_{RC})^2$ can be graphically expressed as  
\begin{eqnarray}\label{Trrho_RC}
\textrm{Tr}(\rho_{RC})^2=\frac{1}{d_A^2}~~\figbox{0.3}{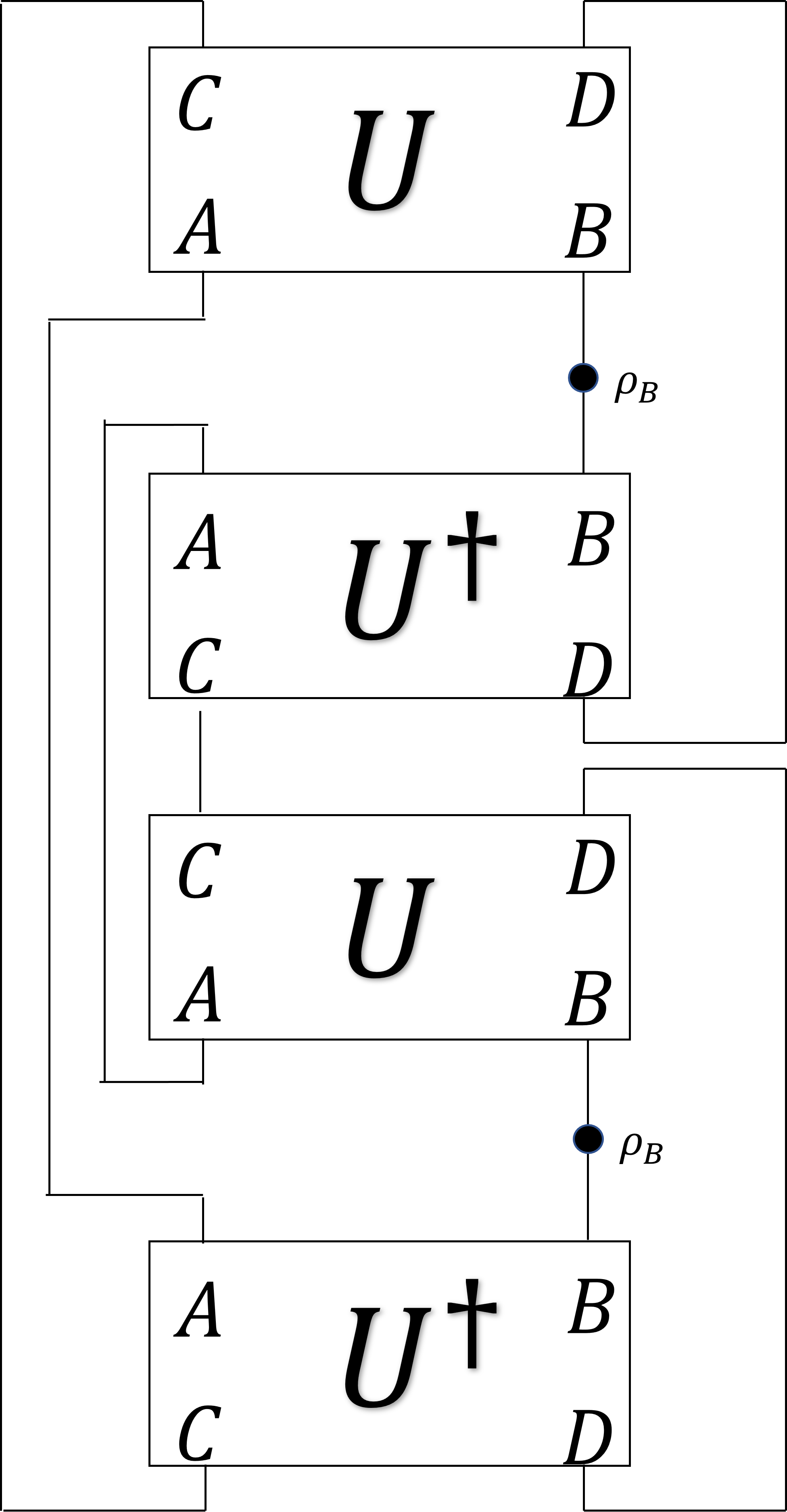} 
\end{eqnarray}
Explicitly, this can be written as 
\begin{eqnarray}
\textrm{Tr}(\rho_{RC})^2&=&\frac{1}{d_A^2}
U_{c_1d_1a_1b_1} (\rho_B)_{b_1b_1'} U^\dagger_{a_2b_1'c_2d_1}
U_{c_2d_2a_2b_2} (\rho_B)_{b_2b_2'} U^\dagger_{a_1b_2'c_1d_2}\nonumber\\
&=&\frac{1}{d_A^2}(\rho_B)_{b_1b_1'}(\rho_B)_{b_2b_2'}
U_{c_1d_1a_1b_1} U_{c_2d_2a_2b_2} 
U^\ast_{c_2d_1a_2b_1'}
  U^\ast_{c_1d_2a_1b_2'}
\end{eqnarray}

In studying the random states, we often want to integrate polynomials of the unitary matrices over the group invariant Haar measure on $U(N)$. The integral over the Haar measure satisfies the following properties \cite{Harlow:2014yka}
\begin{eqnarray}
 \int  dU &=&1,\nonumber\\
 \int dU\ U_{i_1j_1}U^\ast_{i_2j_2}
 &=&\frac{\delta_{i_1i_2}\delta_{j_2j_1}}{d},
 \nonumber\\
 \int dU\  U_{i_1j_1}U_{i_2j_2}U^\ast_{i_3j_3}U^\ast_{i_4j_4}
 &=&\frac{\delta_{i_1i_3}\delta_{i_2i_4}\delta_{j_1j_3}\delta_{j_2j_4}
 + \delta_{i_1i_4}\delta_{i_2i_3}\delta_{j_1j_4}\delta_{j_2j_3}}{d^2-1}
 \nonumber\\
 &&-\frac{\delta_{i_1i_3}\delta_{i_2i_4}\delta_{j_1j_4}\delta_{j_2j_3}
 + \delta_{i_1i_4}\delta_{i_2i_3}\delta_{j_1j_3}\delta_{j_2j_4}}{d(d^2-1)},\nonumber
\end{eqnarray}
where $d$ is the dimension of the Hilbert space that the random unitary matrix $U$ acts on.

Using the above formulas, one can obtain 
\begin{eqnarray}
&&\int dU\textrm{Tr}(\rho_{RC})^2\nonumber\\&=&\frac{1}{d_A^2} (\rho_B)_{b_1b_1'}(\rho_B)_{b_2b_2'}\int dU U_{c_1d_1a_1b_1} U_{c_2d_2a_2b_2} 
U^\ast_{c_2d_1a_2b_1'}   U^\ast_{c_1d_2a_1b_2'}\nonumber\\
 &=&\frac{(\rho_B)_{b_1b_1'}(\rho_B)_{b_2b_2'}}{d_A^2}   \left[ \frac{\left(\delta_{c_1c_2}\delta_{d_1d_1}\delta_{c_2c_1}\delta_{d_2d_2}\delta_{a_1a_2}\delta_{b_1b_1'}\delta_{a_2a_1}\delta_{b_2b_2'}+\delta_{c_1c_1}\delta_{d_1d_2}\delta_{c_2c_2}\delta_{d_2d_1}\delta_{a_1a_1}\delta_{b_1b_2'}\delta_{a_2a_2}\delta_{b_2b_1'} \right)}{d^2-1} \right.\nonumber\\
 &&\left.-\frac{\left(
 \delta_{c_1c_2}\delta_{d_1d_1}\delta_{c_2c_1}\delta_{d_2d_2}\delta_{a_1a_1}\delta_{b_1b_2'}\delta_{a_2a_2}\delta_{b_2b_1'}+\delta_{c_1c_1}\delta_{d_1d_2}\delta_{c_2c_2}\delta_{d_2d_1}\delta_{a_1a_2}\delta_{b_1b_1'}\delta_{a_2a_1}\delta_{b_2b_2'} \right)}{d(d^2-1)}
 \right]\nonumber\\
 &=&\frac{1}{d_A(d^2-1)} 
 \left(dd_D+d^2d_C\langle\rho_B^2\rangle-d d_D \langle\rho_B^2\rangle -d_C \right)\;,
\end{eqnarray}
where we have defined $\langle\hat{O}\rangle=\textrm{Tr}\left(\hat{O}\right)/d_B$ for the operator applied in the Hilbert space of the system $B$ and used the fact that $\langle\rho_B\rangle=1/d_B$. 

Therefore, we have
\begin{eqnarray}
\left(\int dU \|\rho_{RC}-\frac{d_R}{d_C} I_R \otimes I_C \|_1\right)^2 &\leq& d_Rd_C \int dU \textrm{Tr}(\rho_{RC})^2-1
\nonumber\\
&=& \frac{1}{d^2-1} 
 \left(d^2d_C^2\langle\rho_B^2\rangle-d^2 \langle\rho_B^2\rangle -d_C^2+1 \right)\nonumber\\
&=&\frac{\left(d^2\langle\rho_B^2\rangle-1\right)\left(d_C^2-1\right)}{d^2-1} \nonumber\\
&\approx& d_C^2 \langle\rho_B^2\rangle
\nonumber\\
&=&\frac{d^2 \langle\rho_B^2\rangle}{d_D^2}
\;,\label{inequality}
\end{eqnarray}
where we have used the fact that the Hilbert space dimensions is big enough. 

The condition that the reference system $R$ decouples from the remnant black hole $C$ and the information of the message system $A$ can be extracted from the Hawking radiation is then given by 
\begin{eqnarray}\label{decoupcon}
d_D^2\gg d^2 \langle\rho_B^2\rangle\;.
\end{eqnarray}
Under this decoupling condition, the $L_1$ norm between the reduced density matrix $\rho_{RC}$ and the product state of the system $R$ and $C$ averaged over the random unitary matrix $U$ is sufficiently small. This implies that the correlations between the black hole $B$ after swallowing the message system $A$ and the reference system $R$ have been transported to the correlations between the radiation system $B'D$ and the reference system $R$. The information of the message system $A$ can be decoded by the observer Bob who has access to the early radiation system $B'$ and the late radiation system $D$.

In the high temperature limit $\beta_B\rightarrow 0$, one has 
$\langle\rho_B^2\rangle=1/d_B^2$. The Eq. (\ref{inequality}) can be reduced to  
\begin{eqnarray}
\left(\int dU \|\rho_{RC}-\frac{d_R}{d_C} I_R \otimes I_C \|_1\right)^2 \leq \frac{d_C^2}{d_B^2}=\frac{d_A^2}{d_D^2}\;,
\end{eqnarray}
which coincides with the result of Hayden and Preskill \cite{Hayden:2007cs}. In this case, the decoupling condition is given by $d_D^2\gg d_A^2$.

Note that $\rho_B=e^{-\beta_B H_B}/Z_B$, therefore we have 
\begin{eqnarray}
\textrm{Tr}\rho_B^2=\frac{\sum_{n} e^{-2\beta_B E^B_n}}{\left(\sum_{n} e^{-\beta_B E^B_n}\right)^2}\geq \frac{1}{d_B}\;, 
\end{eqnarray}
where the Cauchy–Schwarz inequality is used and the condition for the equality is given by $\beta_B=0$. The above equation implies that $d^2\langle\rho_B^2\rangle\geq d_A^2$. Therefore, the decoupling condition Eq.(\ref{decoupcon}) at finite temperature is stronger than the condition at infinite temperature. This indicates that the observer Bob should collect more radiation in order to decode the message. The reason is that the initially prepared state of the system $B$ and $B'$ for the finite temperature case is the TFD state while the initial state for the infinite temperature is the EPR state. The EPR state is a maximally entangled state while the TFD state is not.

\section{Decoding Hawking radiation at finite temperature}\label{Decoding_HR}

In this section, we consider the decoding strategy of the observer Bob who has collected all of the early radiation $B'$ and the late radiation $D$. Let us assume that he also has the complete knowledge of the unitary dynamics $U$ of the black hole. A similar strategy has been considered in \cite{Cheng:2019yib,Yoshida:2019kyp}, where the out-of-time-order correlators (OTOC) are used to compute the approximate values of the decoding probability and the fidelity. Here, we will employ the random unitary integral method to recompute the the decoding probability and the fidelity. The decoding strategy at the finite temperature can be described as follows, which is analogous to the decoding strategy at the infinite temperature \cite{Yoshida:2017non}.

Firstly, prepare a copy of $|EPR\rangle_{RA}$, which is denoted as $|EPR\rangle_{R'A'}$. Then, apply the random unitary operator $U^\ast$ on the system $B'A'$. The resultant state $|\Psi_{in}\rangle$ is then given by 
\begin{eqnarray}
|\Psi_{in}\rangle&=&(I_{R}\otimes U_{AB}\otimes U^\ast_{B'A'}\otimes I_{R'}) |EPR\rangle_{RA}\otimes |TFD\rangle_{BB'}\otimes |EPR\rangle_{R'A'}\nonumber\\
&=&\figbox{0.3}{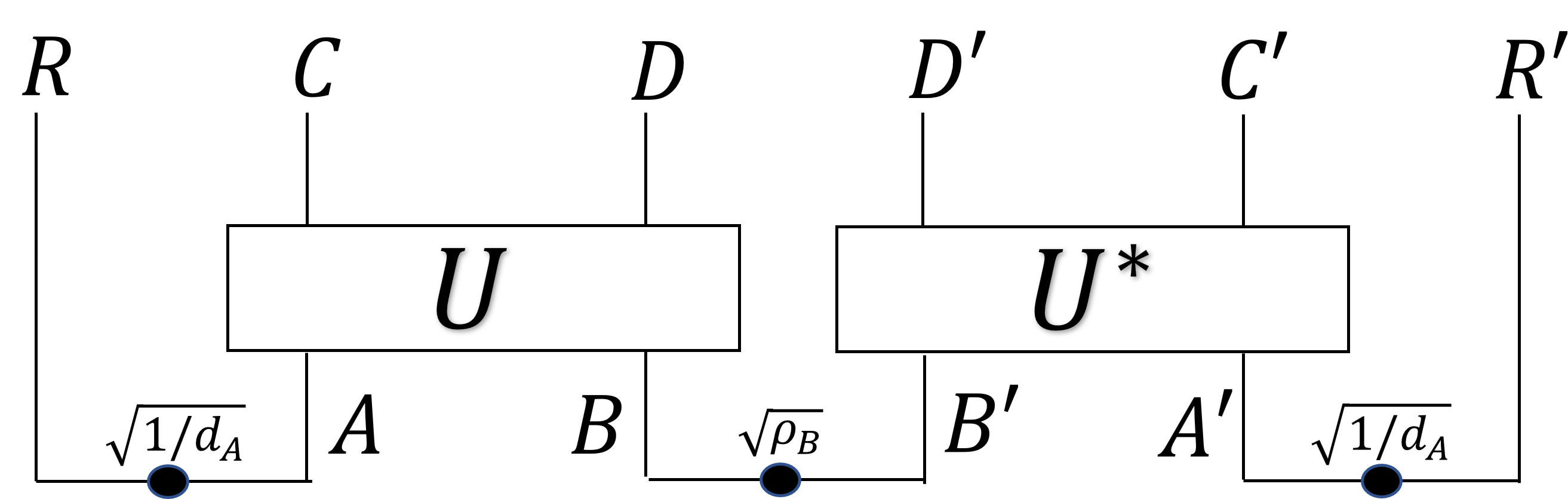} 
\end{eqnarray}
The corresponding density matrix is given by 
\begin{eqnarray}
\rho_{in}=|\Psi_{in}\rangle \langle\Psi_{in}|=\figbox{0.3}{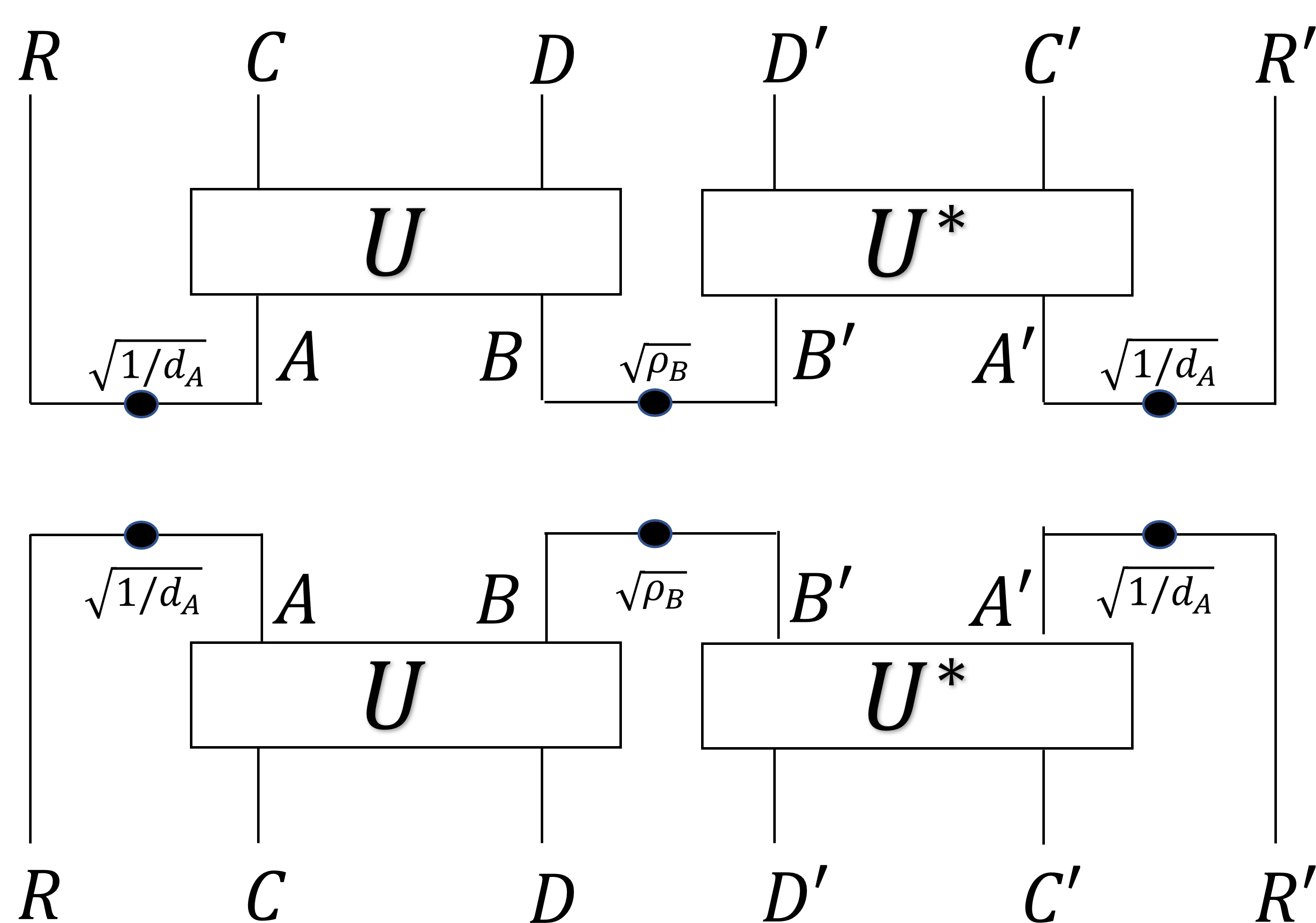} 
\end{eqnarray}

Then, project the state $|\Psi_{in}\rangle$ onto the state $|EPR\rangle_{DD'}$, i.e. act the projecting operator $\Pi_{DD'}=|EPR\rangle_{DD'}\langle EPR|_{DD'}$ to the state $|\Psi_{in}\rangle$. The projecting operation serves to decouple the prepared system $R'$ from the remnant black holes $CC'$ and teleports Alice's quantum state to the prepared system $R'$ owned by Bob \cite{Yoshida:2018vly}. After projecting the state $|\Psi_{in}\rangle$ onto $|EPR\rangle_{DD'}$, the resulted state $|\Psi_{out}\rangle$ is then given by 
\begin{eqnarray}
|\Psi_{out}\rangle&=&\frac{1}{\sqrt{P_{EPR}}}
(I_{RC}\otimes \Pi_{DD'}\otimes I_{R'C'})|\Psi_{in}\rangle\nonumber\\
&=& \frac{1}{\sqrt{P_{EPR}}}~~ 
\figbox{0.3}{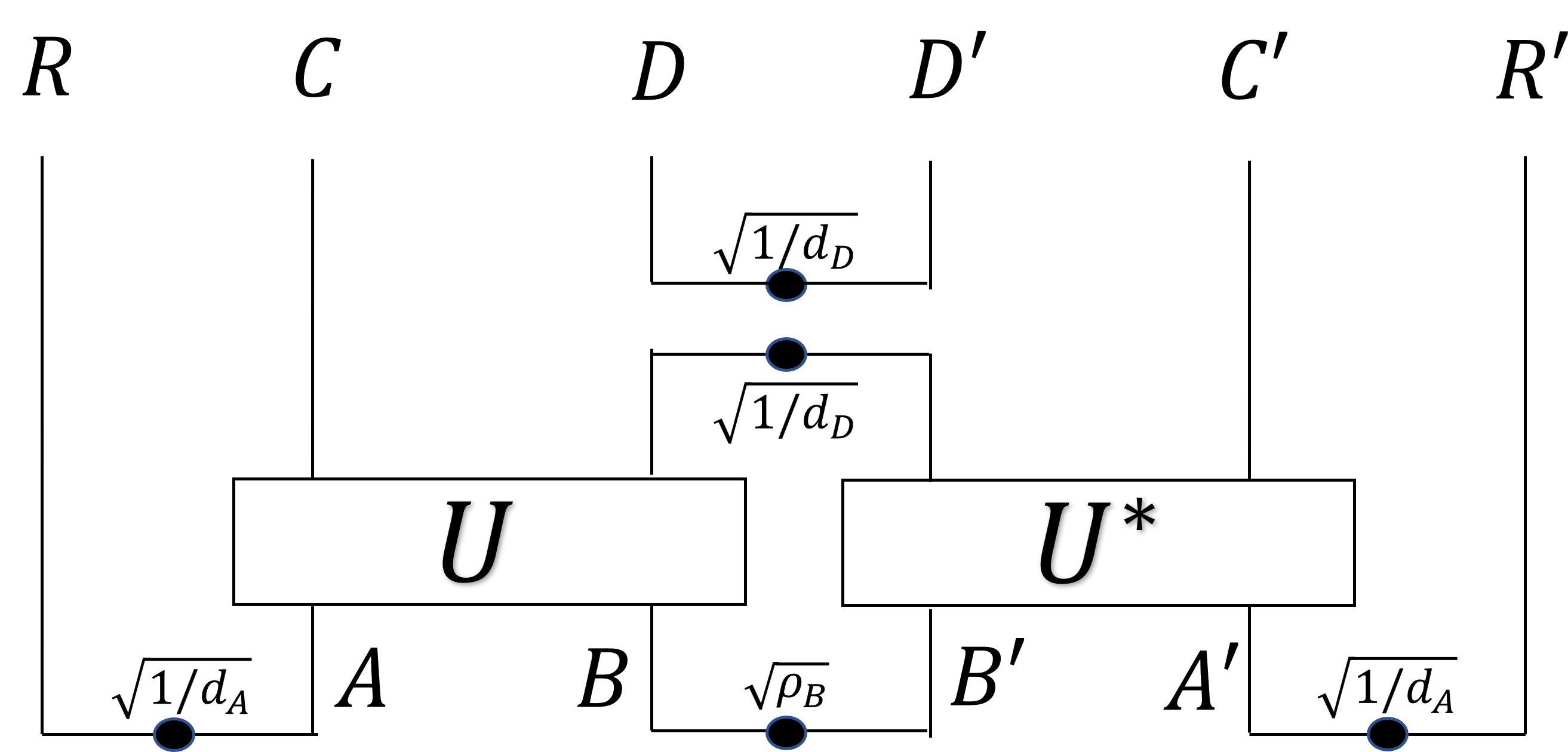}
\end{eqnarray}
where $\frac{1}{\sqrt{P_{EPR}}}$ is the normalization factor and $P_{EPR}$ is the probability of projecting the state $|\Psi_{in}\rangle$ onto the state $|EPR\rangle_{DD'}$.

The probability $P_{EPR}$ is given by 
\begin{eqnarray}
P_{EPR}&=&\textrm{Tr}\left[\Pi_{DD'} \rho_{in}\right]\nonumber\\&=&\frac{1}{d_A^2d_D}~~ \figbox{0.2}{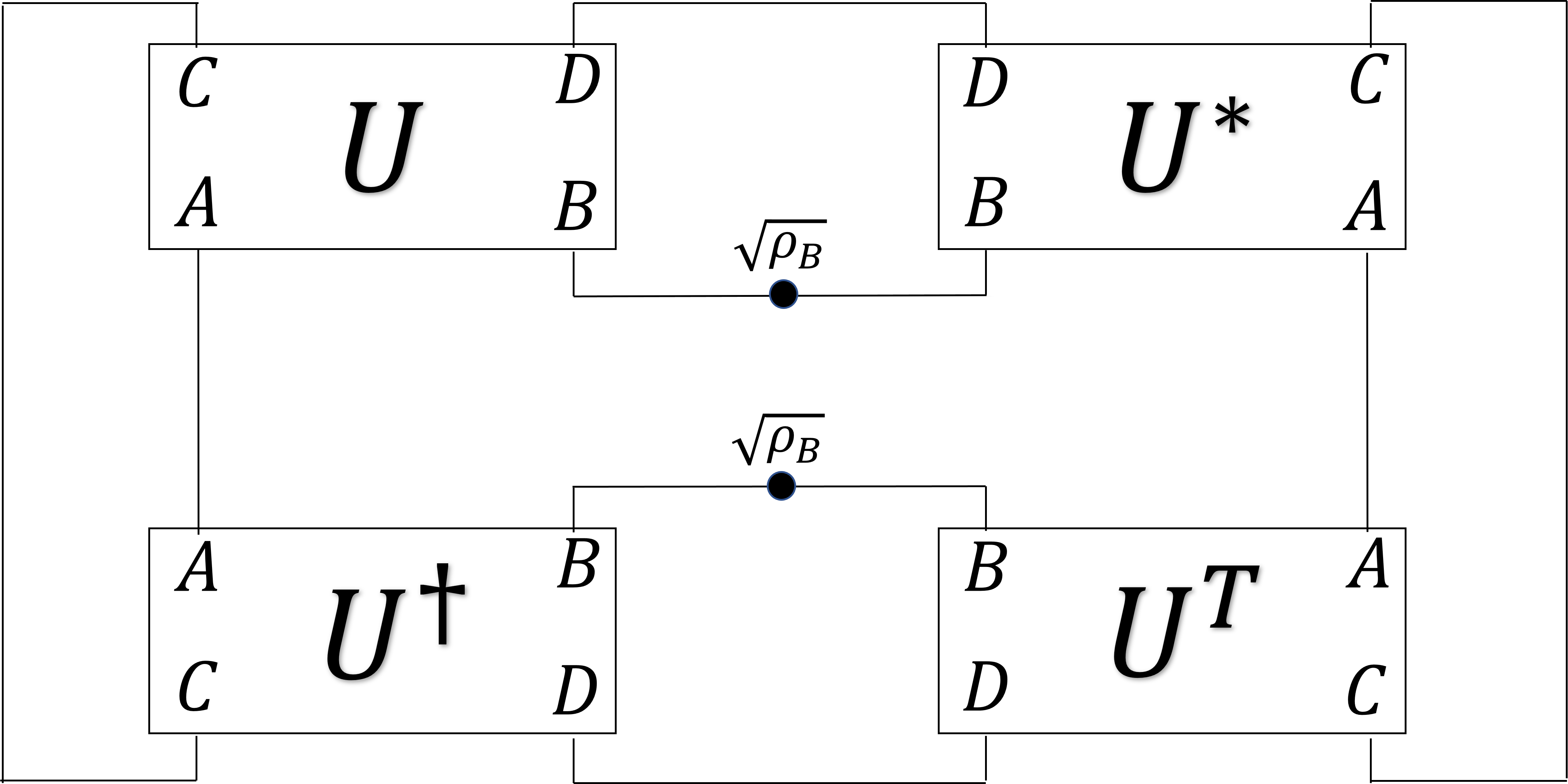}
\end{eqnarray}

After some arrangements, we can get the following equivalent graphical representation of the projecting probability as 
\begin{eqnarray}
P_{EPR}&=&\frac{1}{d_A^2d_D}~~ \figbox{0.2}{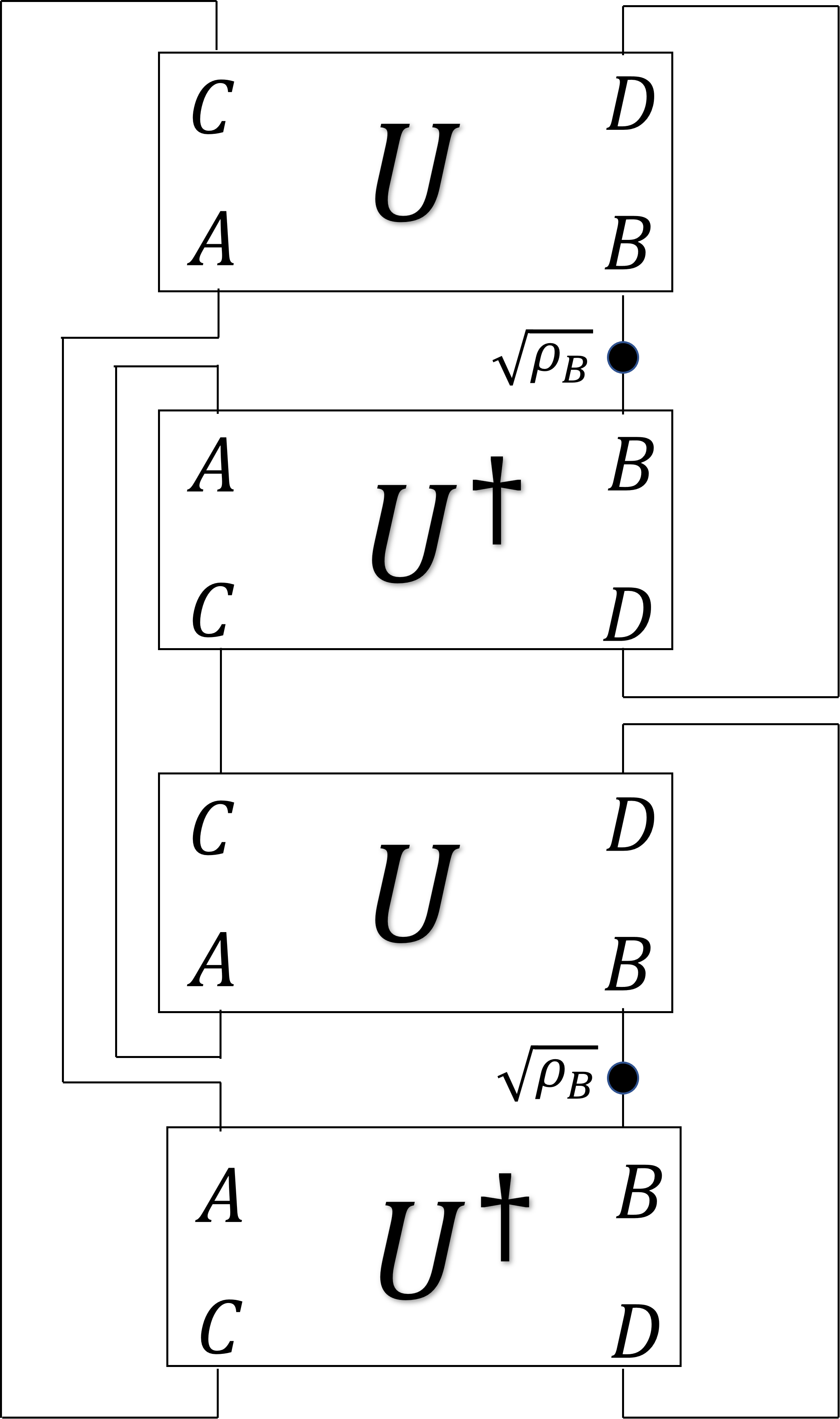} 
\end{eqnarray}
Note that this graphical representation is similar to the graphical representation of the reduced density matrix $\textrm{Tr}\left(\rho_{RC}\right)^2$ in Eq.(\ref{Trrho_RC}) in the last section. However the black dots have different meanings. Explicitly, the probability $P_{EPR}$ can be rewritten as 
\begin{eqnarray}
P_{EPR}=\frac{1}{d_A^2d_D} \left(\sqrt{\rho_B}\right)_{b_1b_1'}\left(\sqrt{\rho_B}\right)_{b_2b_2'}
U_{c_1d_1a_1b_1}  U_{c_2d_2a_2b_2} 
U^\ast_{c_2d_1a_2b_1'}  U^\ast_{c_1d_2a_1b_2'} \;.
\end{eqnarray}
Using the integral formulas of the random unitary matrix, the average value of $P_{EPR}$ can be calculated. After some algebras, the result is given by \begin{eqnarray}\label{P_EPR_ave}
\int dU P_{EPR}&=&\frac{1}{d^2-1}
\left[d_B^3 \langle\sqrt{\rho_B}\rangle^2
+d_C^2-\frac{d_Bd_C^2}{d_A^2} \langle\sqrt{\rho_B}\rangle^2 -1\right]
\nonumber\\
&\approx& \frac{1}{d^2}\left[d_B^3 \langle\sqrt{\rho_B}\rangle^2
+d_C^2\right]\;,
\end{eqnarray}
where the last two terms are ignored in the last step because they are small compared to the first two terms.
This result is consistent with the result in \cite{Cheng:2019yib} calculated by using the OTOC approximation. When $\beta_B\rightarrow 0$, the result also can be reduced to the result in \cite{Yoshida:2017non}. 

Note that $\sqrt{\rho_B}=e^{-\beta_B H_B/2}/\sqrt{Z_B}$, therefore we have 
\begin{eqnarray}
\textrm{Tr}\sqrt{\rho_B}=\frac{\sum_{n} e^{-\beta_B E^B_n/2}}{\sqrt{\sum_{n} e^{-\beta_B E^B_n}}}\leq \sqrt{d_B}\;, 
\end{eqnarray}
where the condition for the equality is $\beta_B=0$. Therefore, we have 
\begin{eqnarray}
\int dU P_{EPR} \leq \frac{1}{d_A^2}+\frac{1}{d_D^2}\approx \frac{1}{d_A^2}\;,
\end{eqnarray}
where  we have used the fact that $d_D^2\gg d_A^2$. This inequality implies that the probability of projecting the state $|\Psi_{in}\rangle$ onto the state $|EPR\rangle_{DD'}$ at the finite temperature is less than that at infinite temperature in \cite{Yoshida:2017non}. 

The fidelity between the out state $|\Psi_{out}\rangle$ and $|EPR\rangle_{RR'}$ quantifies the quality of the Bob's decoding, which is given by 
 \begin{eqnarray}
 F_{EPR}&=&\langle \Psi_{out}| \Pi_{RR'} |\Psi_{out}\rangle
 \nonumber\\
 &=& \frac{1}{P_{EPR} d_A^3 d_D} ~~
 \figbox{0.2}{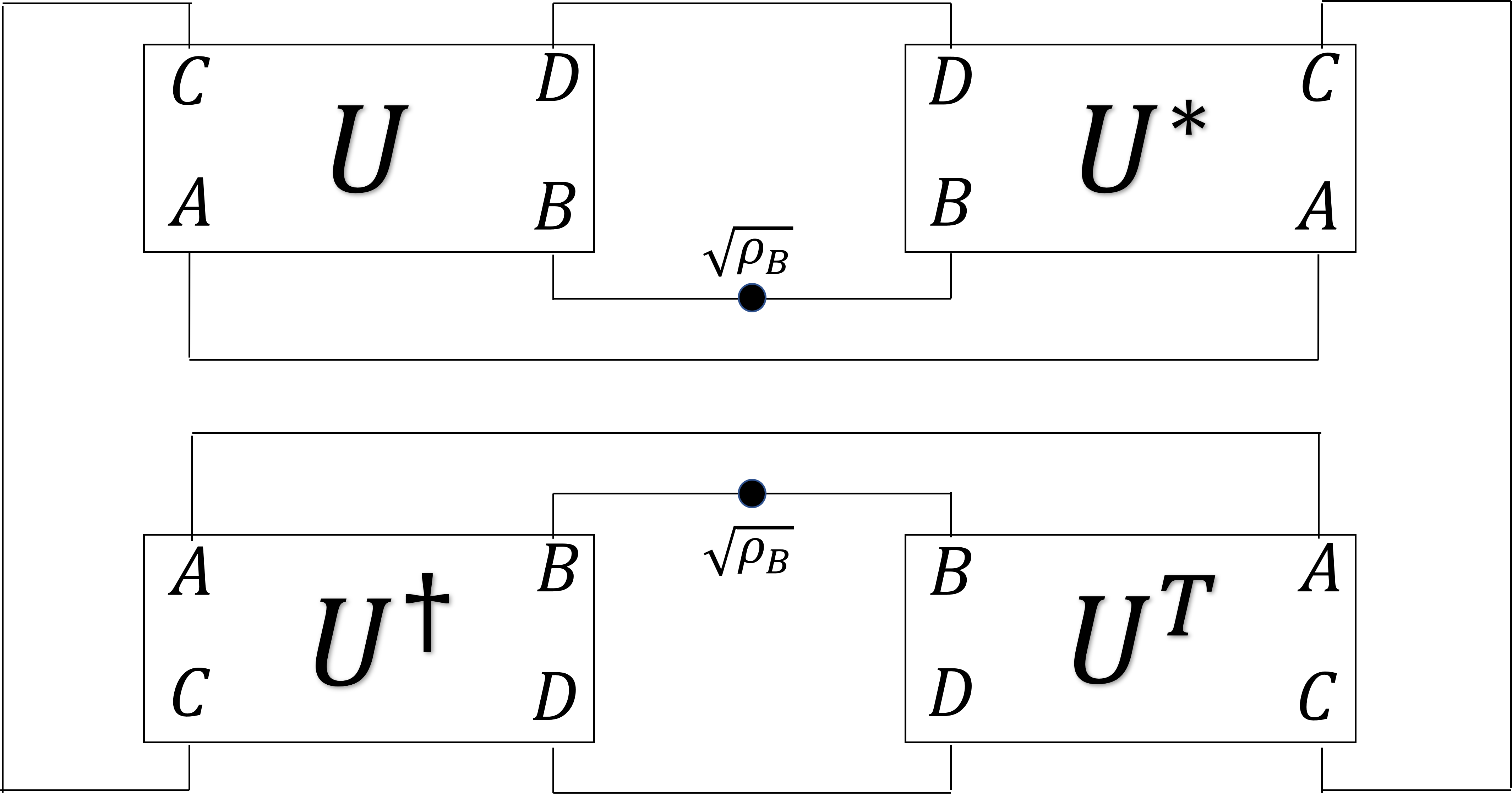}
 \nonumber\\
 &=& \frac{1}{P_{EPR} d_A^3 d_D} ~~
 \figbox{0.2}{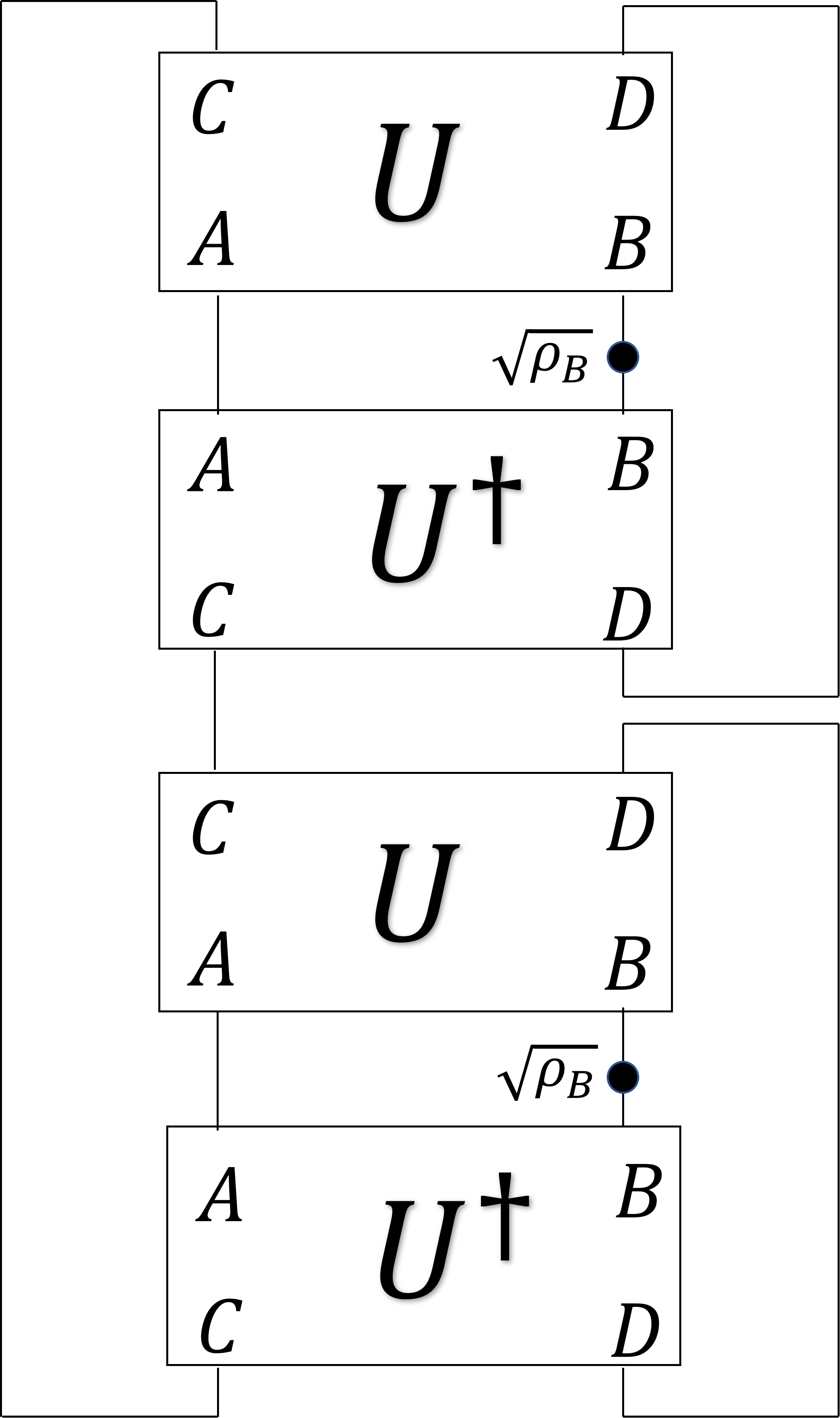}=\frac{\delta}{P_{EPR} d_A^2}
 \end{eqnarray}
 Then, $\delta$ is given explicitly as
 \begin{eqnarray}
 \delta=\frac{1}{d_Ad_D} \left(\sqrt{\rho_B}\right)_{b_1b_1'}\left(\sqrt{\rho_B}\right)_{b_2b_2'}
 U_{c_1d_1a_1b_1}  U_{c_2d_2a_2b_2} 
 U^\ast_{c_2d_1a_1b_1'}  U^\ast_{c_1d_2a_2b_2'} \;.
 \end{eqnarray}
 The average value $\overline{\delta}$ is given by 
 \begin{eqnarray}\label{F_EPR_ave}
 \overline{\delta}&=& \frac{1}{(d^2-1)} 
 \left[d_A^2d_B^3\langle\sqrt{\rho_B}\rangle^2
 +d_C^2-1-d_Bd_C^2 \langle\sqrt{\rho_B}\rangle^2
 \right]\nonumber\\
 &\approx& \frac{1}{d^2} \left[d_A^2d_B^3\langle\sqrt{\rho_B}\rangle^2
 +d_C^2\right]\;.
 \end{eqnarray}
 Therefore, the fidelity is given by
 \begin{eqnarray}
 F_{EPR}&=& \frac{1}{P_{EPR}(d^2-1)} 
 \left[d_B^3\langle\sqrt{\rho_B}\rangle^2
 +\frac{d_C^2}{d_A^2}-\frac{1}{d_A^2}-\frac{d_Bd_C^2}{d_A^2} \langle\sqrt{\rho_B}\rangle^2
 \right]\nonumber\\
 &=& \frac{d_B^3\langle\sqrt{\rho_B}\rangle^2
 +\frac{d_C^2}{d_A^2}}{d_B^3\langle\sqrt{\rho_B}\rangle^2
 +d_C^2}\;.
 \end{eqnarray}
If $d_B^3\langle\sqrt{\rho_B}\rangle^2\gg d_C^2$, the probability $P_{EPR}\rightarrow \frac{d_B^3\langle\sqrt{\rho_B}\rangle^2}{d_2}$, which in turn gives the fidelity approaching $1$. Otherwise, the fidelity is less than $1$. In particular, in another limiting case where $d_B^3\langle\sqrt{\rho_B}\rangle^2\ll \frac{d_C^2}{d_A^2}$, the fidelity approaches $0$. In this limiting case, the information carried by the message system $A$ cannot be extracted by decoding the Hawking radiation.

\section{Decoding Hawking radiation at finite temperature with erasure errors}\label{decoding_HR_ERR}
 
In this section, we generalized the decoding protocol at finite temperature to the case where there are storage errors in the early Hawking radiation that the observer Bob collected. The storage errors are treated as the erasure errors. In this case, the observer Bob cannot access part of the early radiation of the system $B'$ because they are lost or damaged when Bob is collecting the late radiation \cite{Bao:2020zdo}. 
 
The state before the application of the recovery protocol takes the form
\begin{eqnarray}
\figbox{0.3}{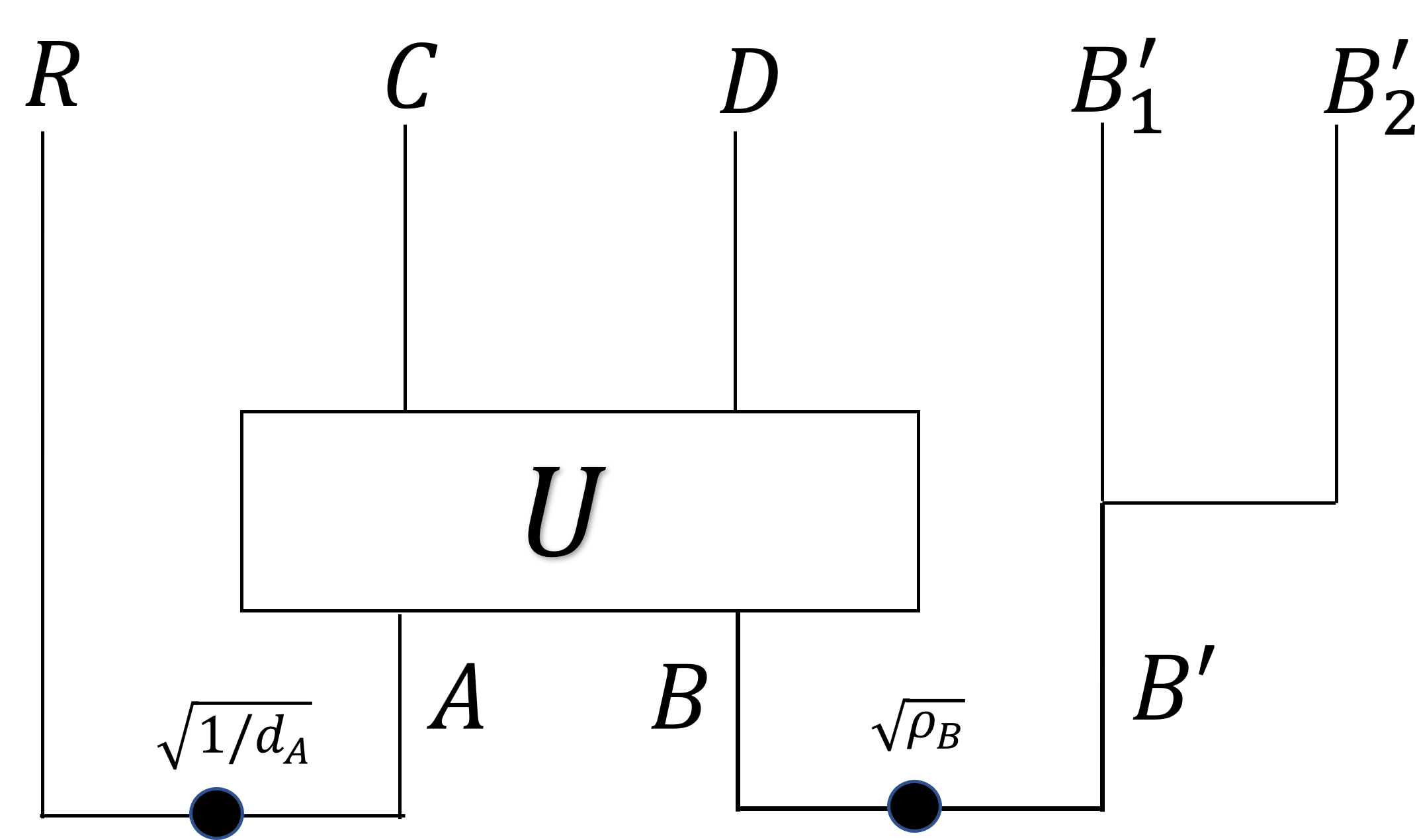} 
\end{eqnarray}
Unlike the graphical representation of $|\Psi_{HP}\rangle$ in Eq. (\ref{psi_HP}), the early radiation system $B'$ splits into two subsystem $B_1'$ and $B_2'$ during the time evolution, where $B_1'$ is the subsystem used to decode the information and $B_2'$ is the lost qubits or erased errors. The erased errors $B_2'$ can be randomly chosen with the probability $p$ from the early radiation system $B'$. This gives us the dimensions of the error system $B_2'$ as $d_{B_2}=(d_B)^p$ and the surviving radiation system $B_1'$ as $d_{B_1}=(d_B)^{1-p}$. 

We can now generalize the decoding protocol at finite temperature to the present case. We assume that the Hilbert space $B'$ can be factorized into the direct product form of the two Hilbert spaces of the subsystems $B_1'$ and $B_2'$, and the density matrix $\rho_B$ can also be factorized as $\rho_B=\rho_{B_1}\otimes \rho_{B_2}$. These conditions reflect that there are no correlations between the early radiations.  

Preparing the EPR state $|EPR\rangle_{R'A'}$, the observer Bod can apply the operator $U^\ast$ on the system $B'A'$. Note that, in order to fill the erased qubits, Bob has to prepare the TFD state of the system $B_2'$. However, the TFD state of the system $B_2'$ will be traced out because Bod is unable to access the erased qubits. The corresponding density matrix of the resultant state $|\Psi_{in}\rangle$ is then given by 
\begin{eqnarray}
\rho_{in}=\figbox{0.3}{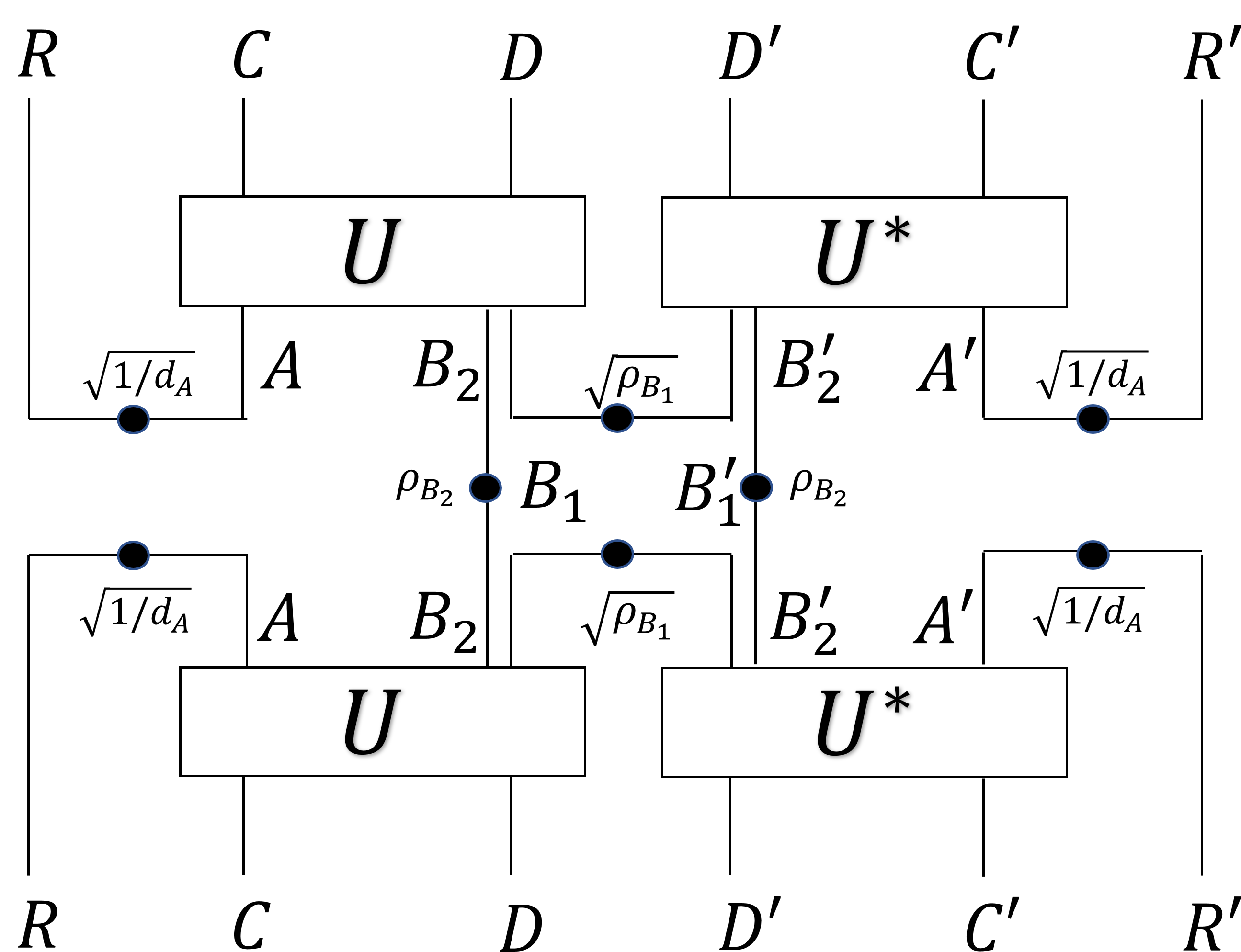} 
\end{eqnarray}

Project the state $|\Psi_{in}\rangle$ onto $|EPR\rangle_{DD'}$, one can obtain the state 
\begin{eqnarray}
\rho_{out}=\frac{\Pi_{DD'}\rho_{in}\Pi_{DD'}}{P_{EPR}},
\end{eqnarray}
where $\Pi_{DD'}=|EPR\rangle_{DD'}\langle EPR|_{DD'}$ is the projector. The probability is given by 
\begin{eqnarray}
P_{EPR}&=&\textrm{Tr}\left[\Pi_{DD'} \rho_{in}\Pi_{DD'}\right]\nonumber\\&=&\frac{1}{d_A^2d_D}~~ \figbox{0.2}{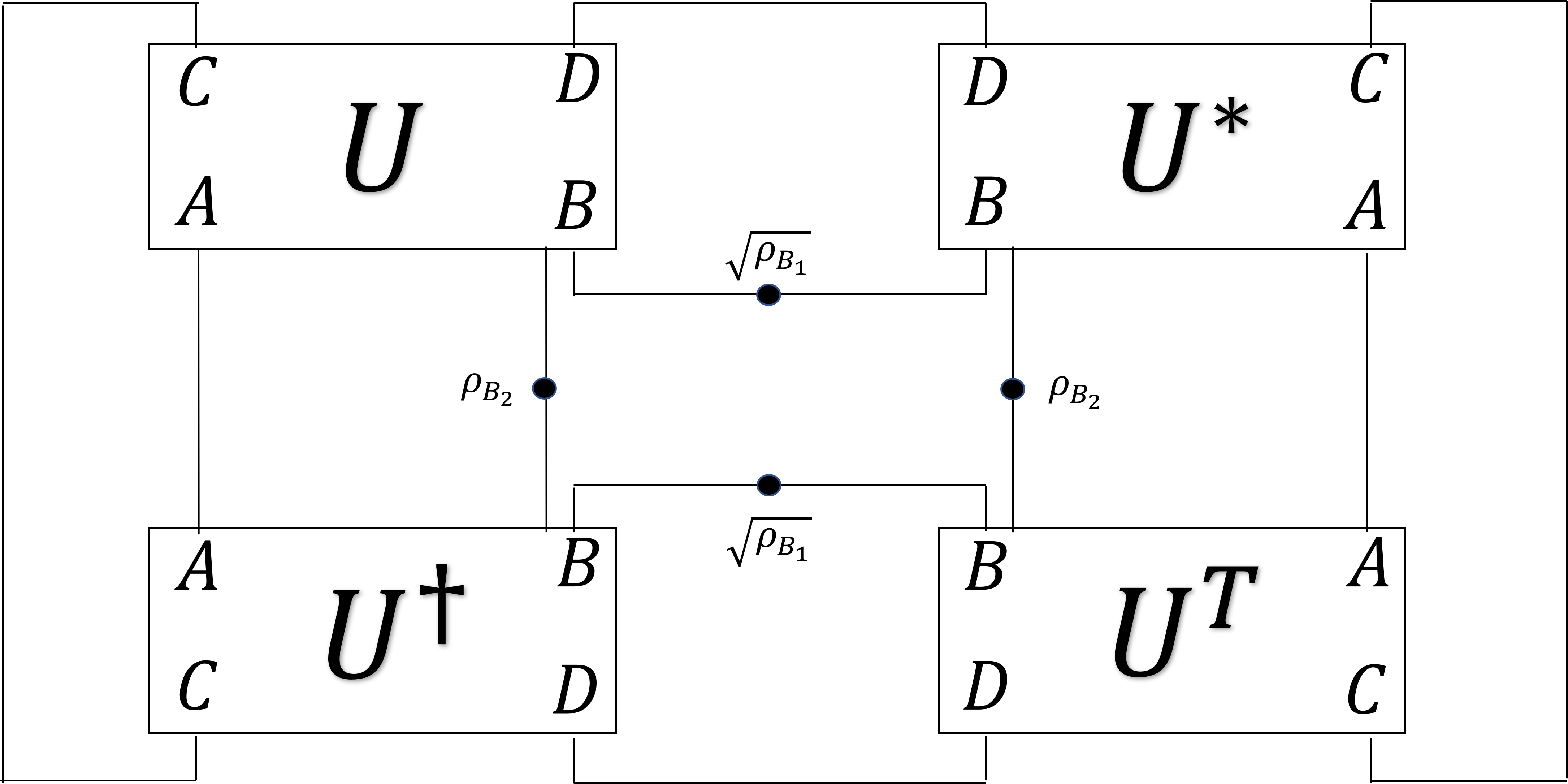}\nonumber\\&=&\frac{1}{d_A^2d_D}~~ \figbox{0.2}{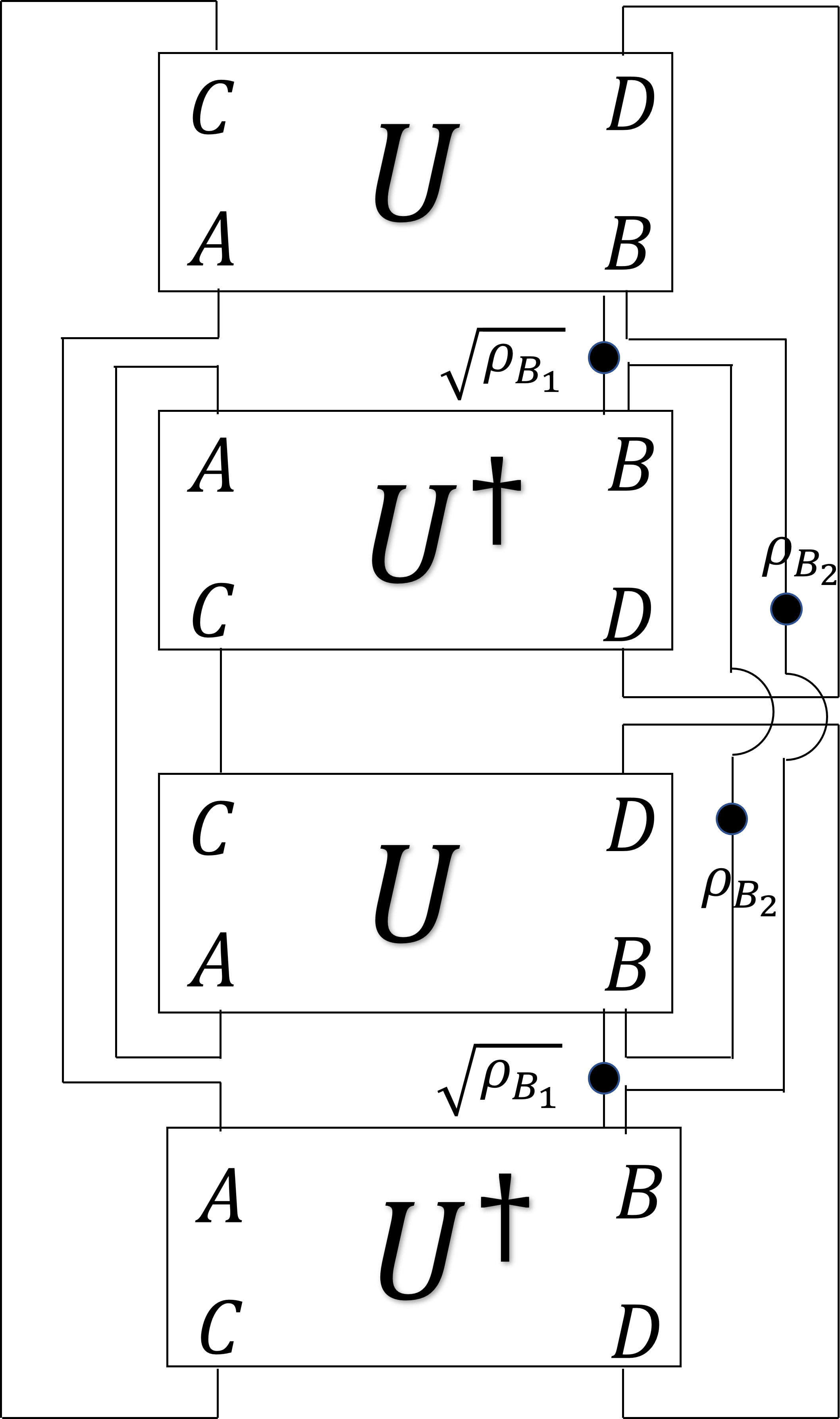} 
\end{eqnarray}
which can be expressed as 
\begin{eqnarray}
P_{EPR}&=&\frac{1}{d_A^2d_D} \left(\sqrt{\rho_{B_1}}\right)_{b_1b_1'}
\left(\sqrt{\rho_{B_1}}\right)_{b_3 b_3'}
\left(\rho_{B_2}\right)_{b_2 b_2'}
\left(\rho_{B_2}\right)_{b_4 b_4'}
\nonumber\\
&&\times
U_{c_1d_1a_1(b_1b_2)}  U_{c_2d_2a_2(b_3b_4')} 
U^\ast_{c_2d_1a_2(b_1'b_4)}  U^\ast_{c_1d_2a_1(b_3'b_2')} \;.
\end{eqnarray}
 
The average value of $P_{EPR}$ is given by 
\begin{eqnarray}\label{P_EPR_err}
\bar{P}_{EPR}&=&\int dU P_{EPR}\nonumber\\
&=&
\frac{1}{d_A^2d_D} \left(\sqrt{\rho_{B_1}}\right)_{b_1b_1'}
\left(\sqrt{\rho_{B_1}}\right)_{b_3 b_3'}
\left(\rho_{B_2}\right)_{b_2 b_2'}
\left(\rho_{B_2}\right)_{b_4 b_4'}\nonumber\\
&&\times 
 \left[ \frac{1}{d^2-1}\left(\delta_{c_1c_2}\delta_{d_1d_1}\delta_{c_2c_1}\delta_{d_2d_2}\delta_{a_1a_2}\delta_{b_1b_1'}\delta_{b_2b_4}\delta_{a_2a_1}\delta_{b_3b_3'}\delta_{b_4'b_2'}
 \right.\right.\nonumber\\
&& \left.
 +\delta_{c_1c_1}\delta_{d_1d_2}\delta_{c_2c_2}\delta_{d_2d_1}\delta_{a_1a_1}\delta_{b_1b_3'}\delta_{b_2b_2'}\delta_{a_2a_2}\delta_{b_3b_1'}\delta_{b_4'b_4} \right) \nonumber\\
 &&-\frac{1}{d(d^2-1)}
 \left(\delta_{c_1c_2}\delta_{d_1d_1}\delta_{c_2c_1}\delta_{d_2d_2}\delta_{a_1a_1}\delta_{b_1b_3'}\delta_{b_2b_2'}\delta_{a_2a_2}\delta_{b_3b_1'}\delta_{b_4'b_4}
 \right.\nonumber\\
&& \left.\left.
 +\delta_{c_1c_1}\delta_{d_1d_2}\delta_{c_2c_2}\delta_{d_2d_1}\delta_{a_1a_2}\delta_{b_1b_1'}\delta_{b_2b_4}\delta_{a_2a_1}\delta_{b_3b_3'}\delta_{b_4'b_2'} \right)\right]\nonumber\\
&=&\frac{1}{d^2-1}
\left[d_B^{3-p} \langle\sqrt{\rho_1}\rangle^2 \langle\rho_2^2\rangle
+d_C^2-\frac{d_B^{1-p}d_C^2}{d_A^2} \langle\sqrt{\rho_1}\rangle^2 \langle\rho_2^2\rangle -1\right]
\nonumber\\
&\approx&\frac{1}{d^2}
\left[d_B^{3-p} \langle\sqrt{\rho_1}\rangle^2 \langle\rho_2^2\rangle +d_C^2 \right]\;,
\end{eqnarray}
where in the last step the last two terms are ignored because they are small compared with the first two terms. It is clear that the projecting probability depends on the expected values of $\sqrt{\rho_1}$ and $\rho_2^2$. When $p=0$, the projecting probability without errors, which is given by Eq.(\ref{P_EPR_ave}) can be recovered from the above equation. If we don't consider the effect of the finite temperature, by replacing $\rho_{B_1}$ and $\rho_{B_2}$ with $1/d_{B_1}$ and $1/d_{B_2}$ respectively, the result can be reduced to the projecting probability at infinite temperature obtained in \cite{Bao:2020zdo}.

The fidelity between the out state $|\Psi_{out}\rangle$ and $|EPR\rangle_{RR'}$ quantifies the quality of the decoding, which is given by 
 \begin{eqnarray}
 F_{EPR}&=&\frac{1}{P_{EPR} d_A^3 d_D} ~~
 \figbox{0.2}{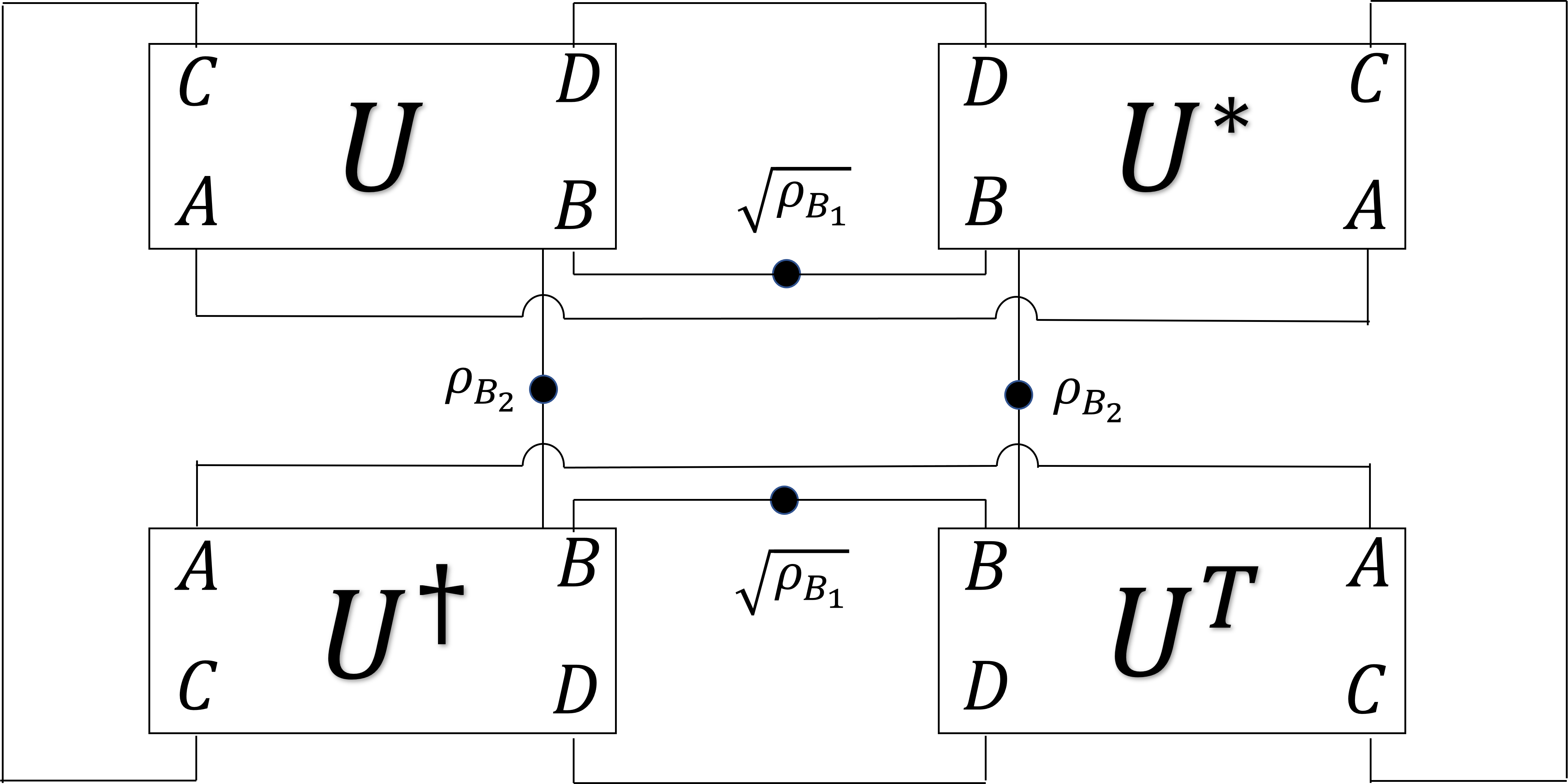}\nonumber\\
 &=&\frac{1}{P_{EPR} d_A^3 d_D} ~~
 \figbox{0.2}{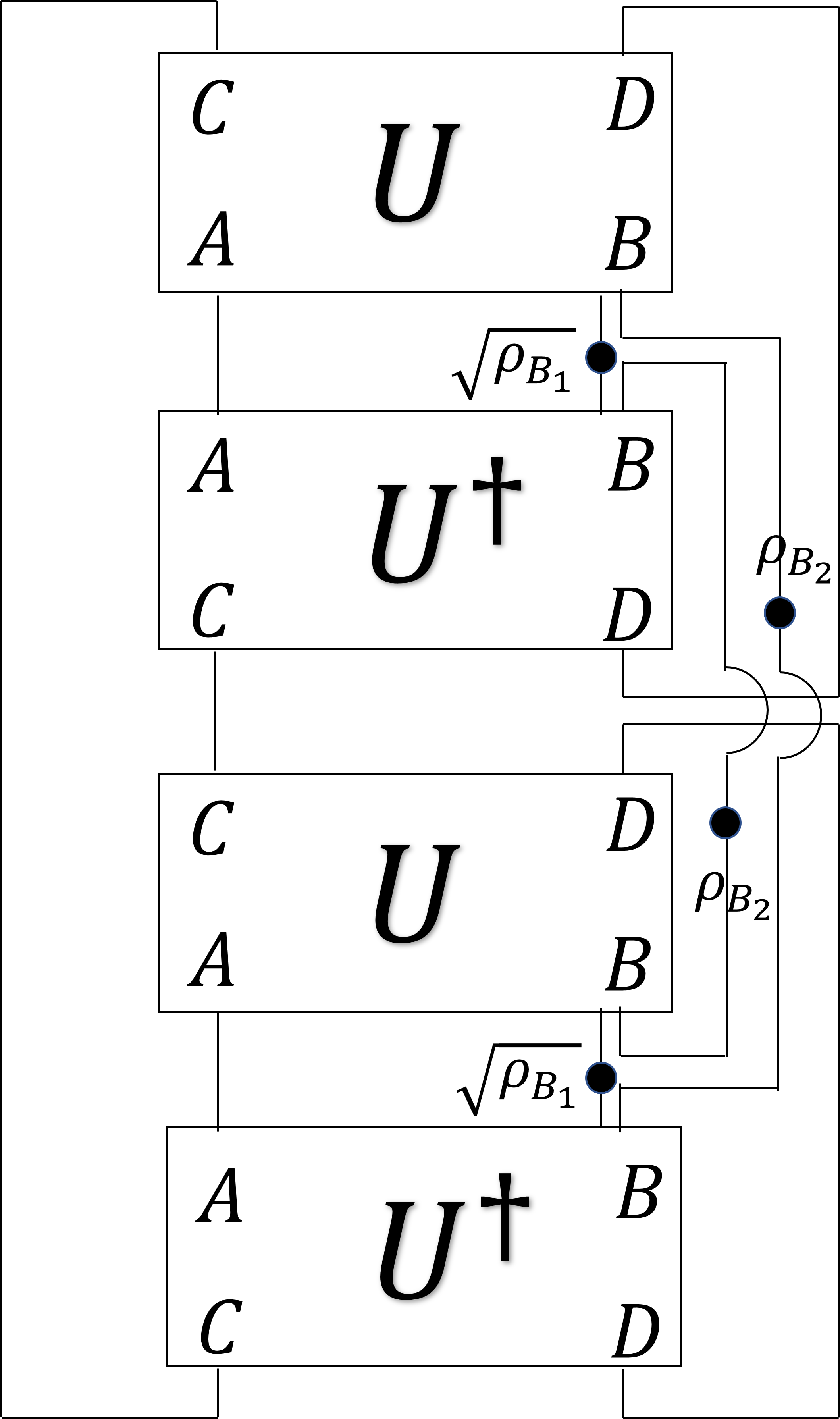}
 =\frac{\delta}{P_{EPR} d_A^2}\;,
 \end{eqnarray}
 where $\delta$ can be expressed as 
 \begin{eqnarray}
 \delta&=&\frac{1}{d_Ad_D} \left(\sqrt{\rho_{B_1}}\right)_{b_1b_1'}
 \left(\sqrt{\rho_{B_1}}\right)_{b_3b_3'}
 \left(\rho_{B_2}\right)_{b_2b_2'}
 \left(\rho_{B_2}\right)_{b_4b_4'}\nonumber\\
 &&\times
 U_{c_1d_1a_1(b_1b_2)}  U_{c_2d_2a_2(b_3b_4')} 
 U^\ast_{c_2d_1a_1(b_1'b_4)}  U^\ast_{c_1d_2a_2(b_3'b_2')} \;.
 \end{eqnarray}
 The average value $\bar{\delta}$ is given by 
 \begin{eqnarray}
 \bar{\delta}&=&\int dU \delta \nonumber\\
 &=& \frac{1}{d^2-1} 
 \left[d^2d_B^{1-p} \langle\sqrt{\rho_1}\rangle^2 \langle\rho_2^2\rangle
+d_C^2-d_C^2 d_B^{1-p} \langle\sqrt{\rho_1}\rangle^2 \langle\rho_2^2\rangle -1\right]
\nonumber\\
&\approx& \frac{1}{d^2} 
 \left[d^2d_B^{1-p} \langle\sqrt{\rho_1}\rangle^2 \langle\rho_2^2\rangle
+d_C^2\right]\;.
 \end{eqnarray}
 
Therefore, the fidelity of the decoding is then given by 
 \begin{eqnarray}
 F_{EPR}&=&\frac{\bar{\delta}}{\bar{P}_{EPR} d_A^2}
 \nonumber\\
 &=&\frac{d^2d_B^{1-p} \langle\sqrt{\rho_1}\rangle^2 \langle\rho_2^2\rangle +d_C^2-d_C^2 d_B^{1-p} \langle\sqrt{\rho_1}\rangle^2 \langle\rho_2^2\rangle -1}{d^2d_B^{1-p} \langle\sqrt{\rho_1}\rangle^2 \langle\rho_2^2\rangle +d_A^2d_C^2-d_B^{1-p}d_C^2 \langle\sqrt{\rho_1}\rangle^2 \langle\rho_2^2\rangle -d_A^2}\nonumber\\
 &\approx&\frac{d_B^{3-p} \langle\sqrt{\rho_1}\rangle^2 \langle\rho_2^2\rangle +\frac{d_C^2}{d_A^2}}{d_B^{3-p} \langle\sqrt{\rho_1}\rangle^2 \langle\rho_2^2\rangle +d_C^2}\;.
 \end{eqnarray}

This result indicates that the fidelity depends on the value of $\langle\sqrt{\rho_1}\rangle^2 \langle\rho_2^2\rangle$. In general, the fidelity is less than $1$. If $d_B^{3-p} \langle\sqrt{\rho_1}\rangle^2 \langle\rho_2^2\rangle\gg d_C^2$, the fidelity approaches unity. The information can be extracted by using the decoding process successfully. If $d_B^{3-p} \langle\sqrt{\rho_1}\rangle^2 \langle\rho_2^2\rangle\ll \frac{d_C^2}{d_A^2}$, the fidelity approaches $0$. In this limiting case, the information carried by the message system $A$ cannot be extracted by decoding the Hawking radiation. In addition, in the infinite temperature limit,  our results can be reduced to the ones obtained in \cite{Bao:2020zdo}.

\section{Decoding Hawking radiation at finite temperature with the depolarizing channel}\label{Decoding_HR_dep}

In this section, we study the effect of decoherence on the decoding process. The decoherence and noise effects on the random unitary dynamics and on the storage of Hawking radiation are considered in \cite{Yoshida:2018vly} and \cite{Bao:2020zdo}, respectively. It is well known that the quantum depolarizing channel is an important type of quantum noise in quantum systems. For a $d$-dimensional quantum system, the depolarizing channel replaces the quantum system with the completely mixed state $I/d$ with probability $p$, and leaves the state untouched otherwise. The corresponding quantum operation \cite{Nielsen:2000} is given by $\mathcal{Q}(\rho)=(1-p)\rho+p \frac{I}{d}$.

For our purpose, we can generalize the quantum depolarizing channel to the finite temperature case by replacing the the completely mixed state $I/d$ with the Gibbs state at finite temperature. Therefore, the depolarizing channel at finite temperature, which maps a state onto a linear combination of itself and the maximally mixed state, can be expressed as 
\begin{eqnarray}
\mathcal{Q}(\rho)=(1-p)\rho+p \rho^{G} \textrm{Tr}\rho\;,
\end{eqnarray}
where $p$ is the probability of depolarization and $\rho^{G}=\frac{1}{Z}\sum_{n}e^{-\beta E_n} |n\rangle\langle n|$ is the Gibbs state.

We consider that the depolarization channel acts on the early radiation system $B'$. However, it can be easily checked that 
\begin{eqnarray}\label{Q_eq}
&&\mathcal{Q}_{B'}(|TFD\rangle\langle TFD|_{BB'})\nonumber\\&=&
\frac{1}{Z} \sum_{n,m} e^{-\beta E_n/2}e^{-\beta E_m/2} 
|n_{B}\rangle \langle m_B| \otimes \mathcal{Q}_{B'} |n_{B'}\rangle
 \langle m_{B'}|\nonumber\\
 &=& (1-p) |TFD\rangle\langle TFD|_{BB'}+
 \frac{p}{Z} \sum_{n,m} e^{-\beta E_n/2}e^{-\beta E_m/2} \textrm{Tr}\left[ |n_{B'}\rangle \langle m_{B'}|\right]
|n_{B}\rangle \langle m_B| \otimes \rho^{G}_{B'} \nonumber\\
&=& (1-p) |TFD\rangle\langle TFD|_{BB'}+
  \frac{p}{Z} \sum_{n} e^{-\beta E_n} |n_{B}\rangle \langle n_B| \otimes \rho^{G}_{B'}\nonumber\\
 &=& (1-p) |TFD\rangle\langle TFD|_{BB'}+
 p \rho^{G}_{B} \otimes \rho^{G}_{B'}\nonumber\\
 &=&\mathcal{Q}_{BB'}(|TFD\rangle\langle TFD|_{BB'})\nonumber\\
 &=&(1-p)~~~\figbox{0.3}{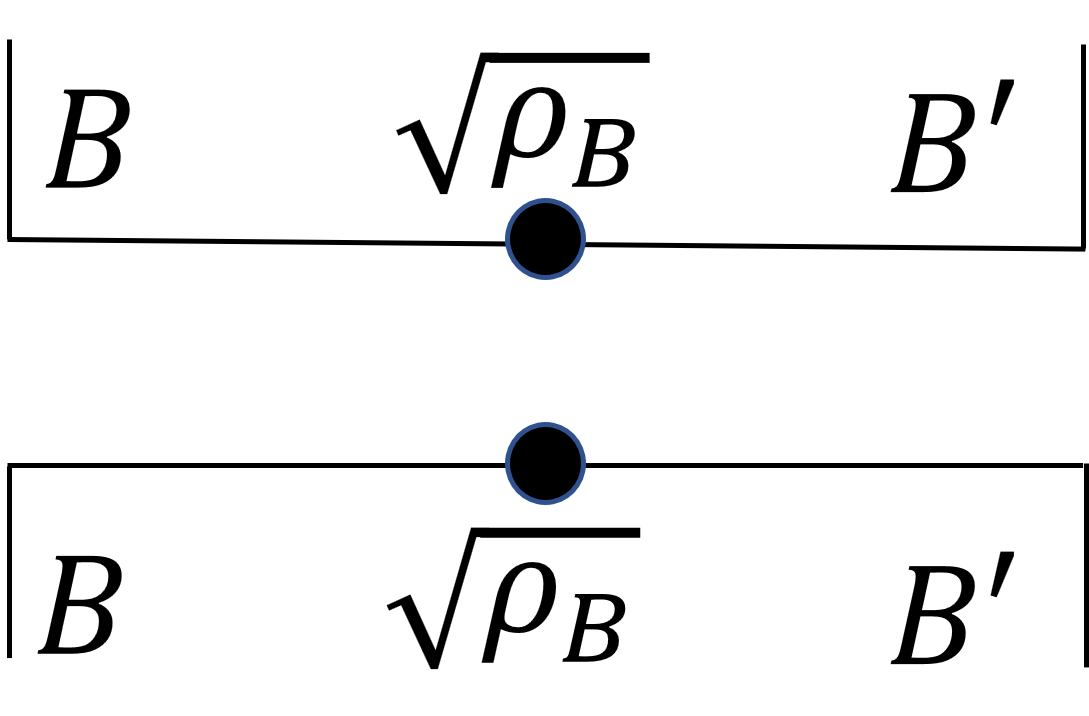}+p ~~~\figbox{0.3}{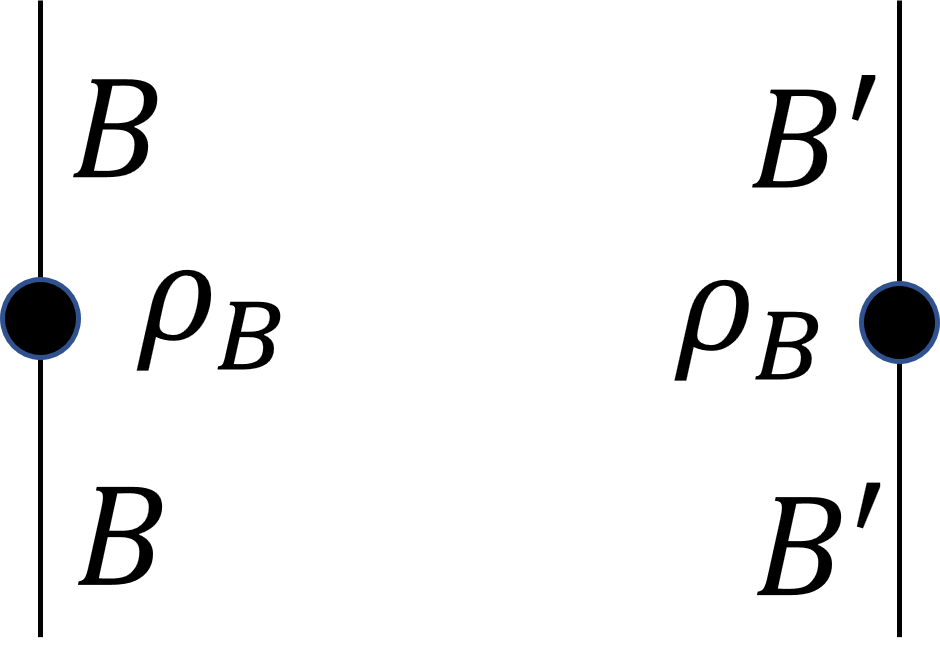}\;. 
\end{eqnarray}
This implies that the quantum depolarization channel can be understood as acting on the state $|TFD\rangle\langle TFD|_{BB'}$ of the system $B$ and $B'$.  

Following the decoding strategy, the observer Bob prepares the EPR state $|EPR\rangle_{R'A'}$, and applies the unitary matrix $U^\ast$ on $B'A'$. The density matrix of the state $\Psi_{in}$ is then given by 
\begin{eqnarray}
\rho_{in}&=&
(I_R\otimes U_{AB}\otimes U^\ast_{B'A'}\otimes I_{R'})\mathcal{Q}_{BB'} (|EPR\rangle\langle EPR|_{RA}\otimes |TFD\rangle\langle TFD|_{BB'}\otimes 
|EPR\rangle\langle EPR|_{RA})\nonumber\\
&&
\times (I_R\otimes U^\dagger_{AB}\otimes U^T_{B'A'}\otimes I_{R'})
\nonumber\\
&=&
\figbox{0.3}{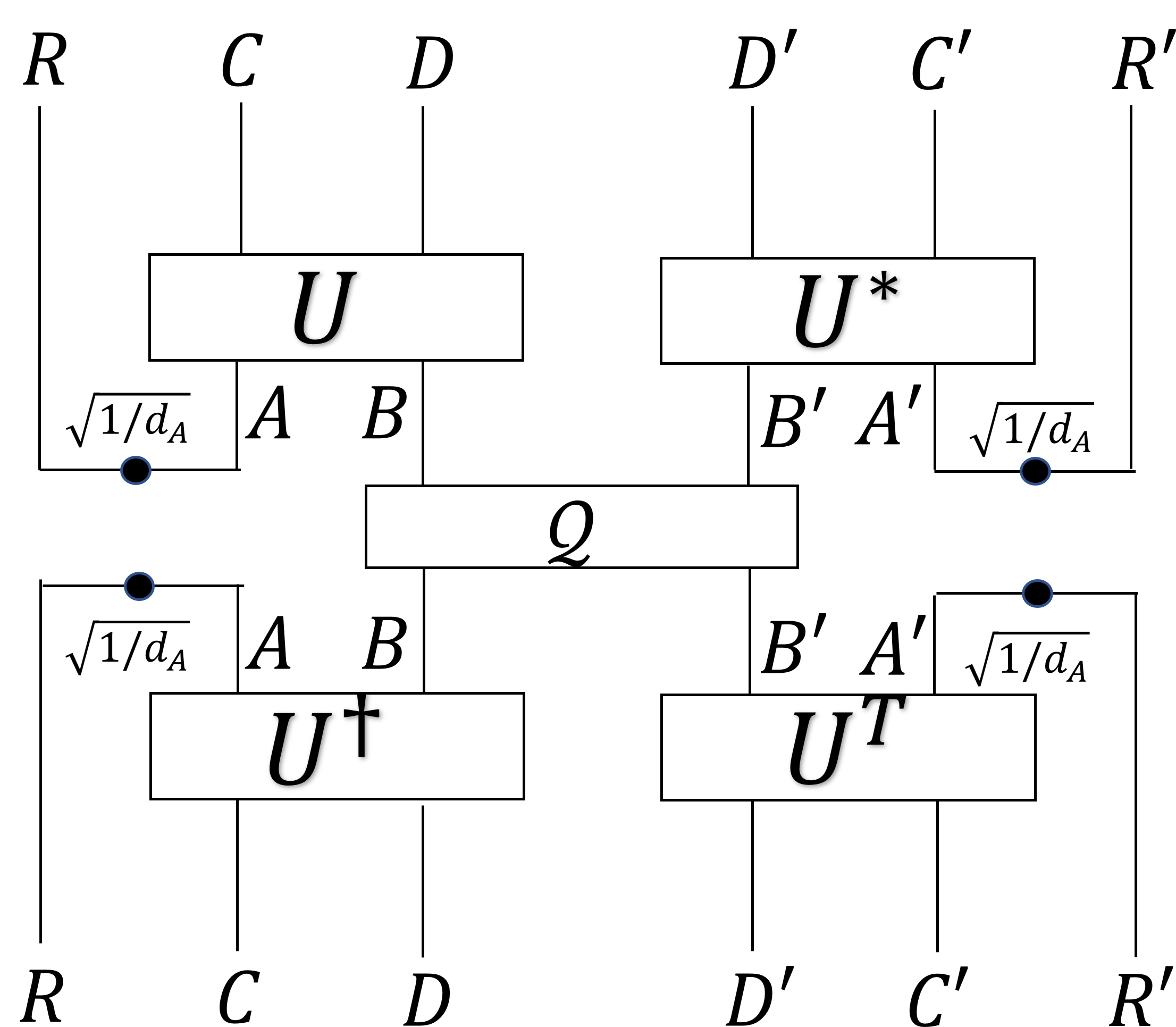} 
\end{eqnarray}

Project the in state $|\Psi_{in}\rangle$ onto $|EPR\rangle_{DD'}$, one can obtain the out state $\rho_{out}=\frac{\Pi_{DD'}\rho_{in}\Pi_{DD'}}{P_{EPR}}$, where the projecting probability $P_{EPR}$ is given by
\begin{eqnarray}
P_{EPR}&=&\textrm{Tr}\left[\Pi_{DD'} \rho_{in}\Pi_{DD'}\right]\nonumber\\&=&\frac{1}{d_A^2 d_D} \figbox{0.2}{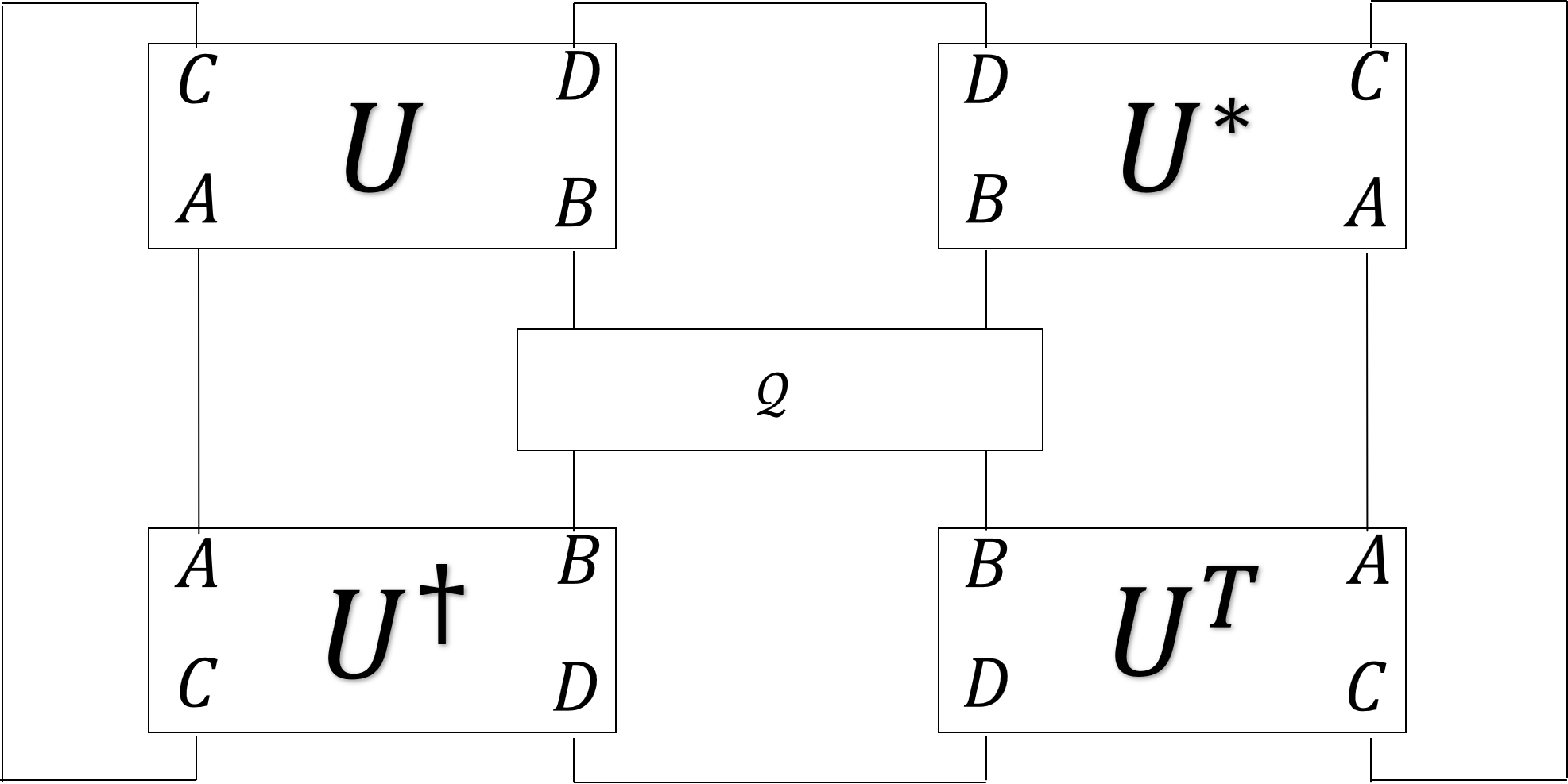}\;.
\end{eqnarray}
According to Eq.(\ref{Q_eq}), $P_{EPR}$ can be further expressed as 
\begin{eqnarray}
P_{EPR}&=&\frac{1-p}{d_A^2 d_D}~~~\figbox{0.15}{PEPR.png}+\frac{p}{d_A^2 d_D}~~~\figbox{0.15}{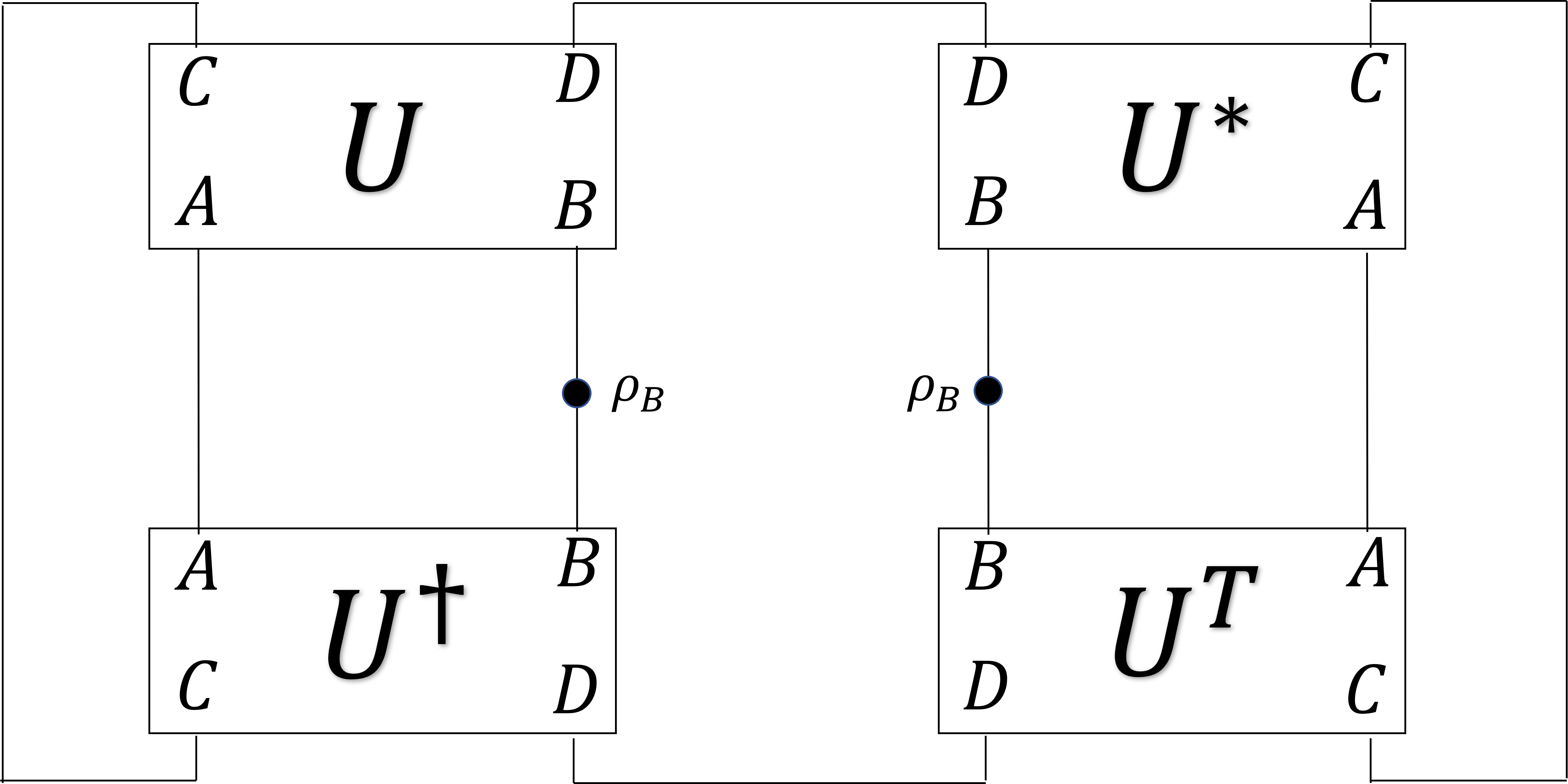}\nonumber\\
&=&\frac{1-p}{d_A^2 d_D}~~~\figbox{0.2}{PEPR_cal.png}+\frac{p}{d_A^2 d_D}~~~\figbox{0.2}{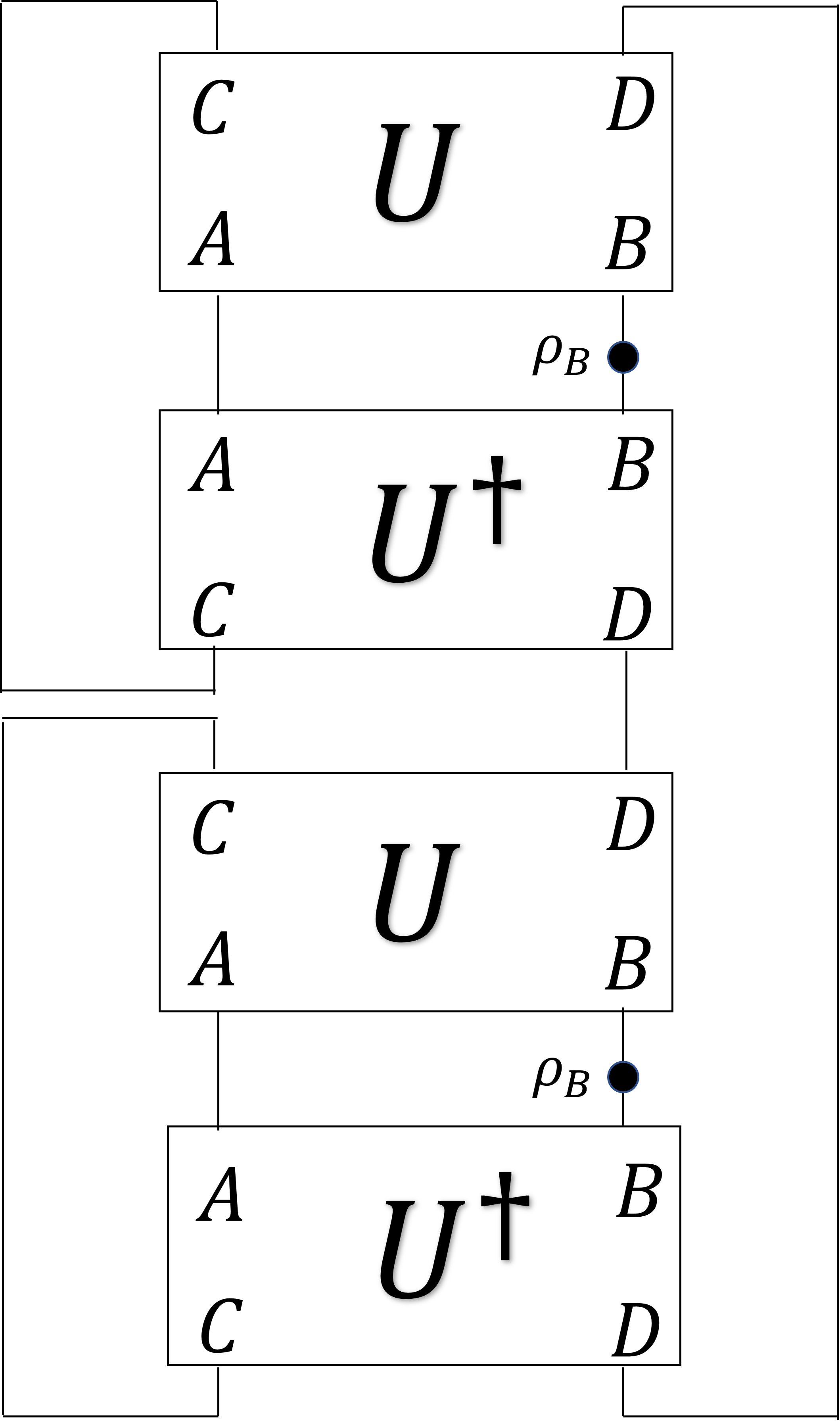}
\end{eqnarray}
The first graph has been calculated and the result is given by Eq.(\ref{P_EPR_ave}). The average value of the second graph can be explicitly calculated as  
\begin{eqnarray}
&&\int dU \left(\rho_{B}\right)_{b_1b_1'}
 \left(\rho_{B}\right)_{b_2b_2'}
  U_{c_1d_1a_1b_1}  U_{c_2d_2a_2b_2} 
 U^\ast_{c_2d_1a_2b_2'}  U^\ast_{c_1d_2a_1b_1'}\nonumber\\
&=&
\frac{d_A^2 d_D}{(d^2-1)} 
 \left[d_B^2\langle\rho_B^2\rangle
 +d_C^2-\frac{d_C^2}{d_A^2} \langle\rho_B^2\rangle-1
 \right]
\end{eqnarray}
Therefore, the average value of $P_{EPR}$ is given by 
\begin{eqnarray}\label{P_EPR_ave_de}
\bar{P}_{EPR}=\int dU P_{EPR}=\frac{1}{d^2-1}
\left[\frac{d^2-d_C^2}{d_A^2}\left((1-p)d_B \langle\sqrt{\rho_B}\rangle^2
+p \langle\rho_B^2\rangle\right)+d_C^2-1\right]
\end{eqnarray}
Once again, this result together with the projecting probabilities given in Eq.(\ref{P_EPR_ave}) and Eq.(\ref{P_EPR_err}) indicate that the projecting probability at the finite temperature depends on the expected value of the thermal density matrix. When $p=0$, the projecting probability in Eq.(\ref{P_EPR_ave_de}) reproduces the probability without errors or depolarizing channel given in Eq.(\ref{P_EPR_ave}). In addition, if the effect of the finite temperature is not considered, the result can be reduced to the projecting probability at infinite temperature obtained in \cite{Bao:2020zdo}.

The fidelity between the out state $|\Psi_{out}\rangle$ and $|EPR\rangle_{RR'}$
 quantifies the quality of decoding, which is given by 
 \begin{eqnarray}
 F_{EPR}&=&\frac{1}{P_{EPR} d_A^3 d_D} ~~
 \figbox{0.2}{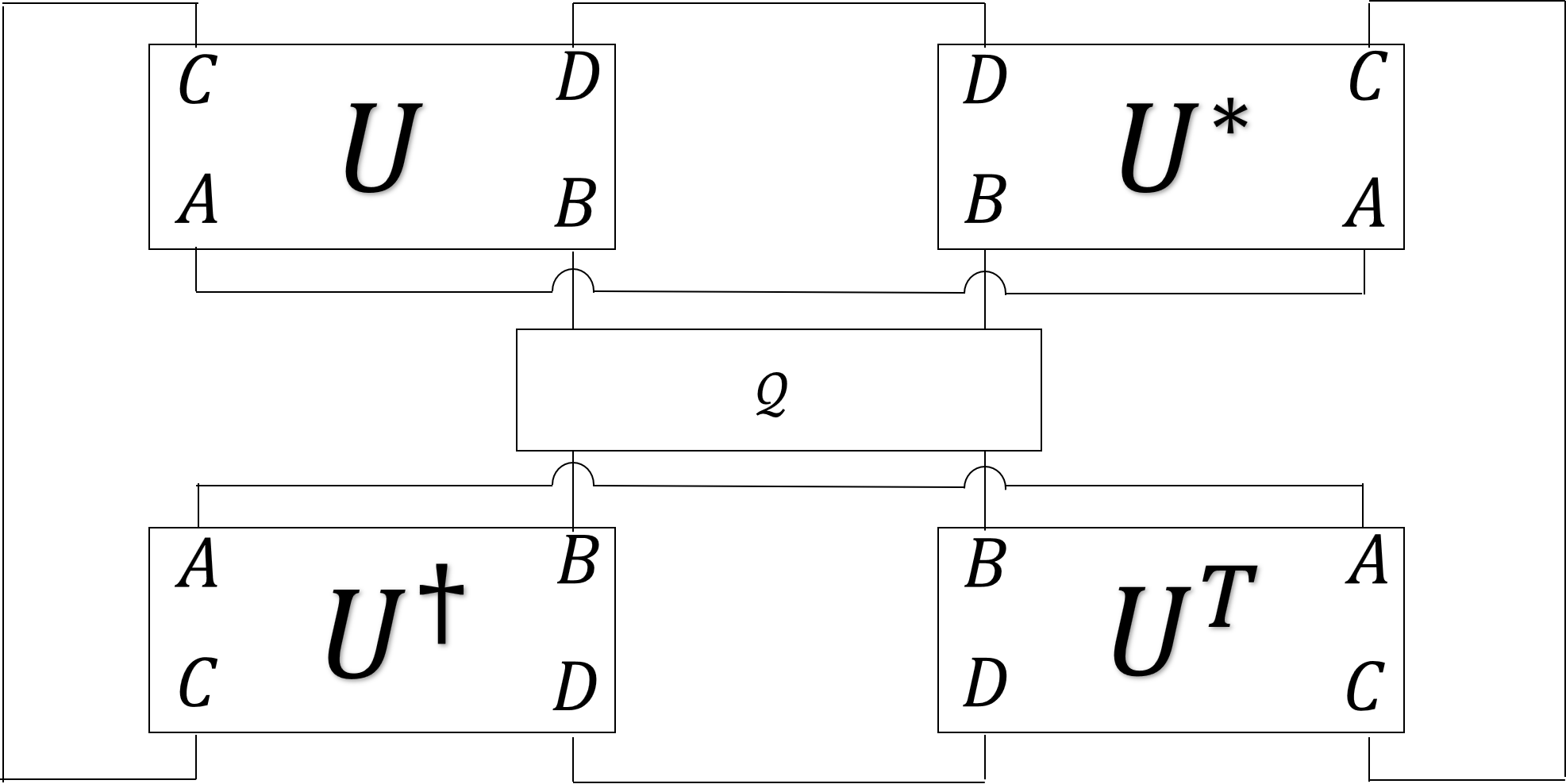}
 \end{eqnarray}
 According to Eq.(\ref{Q_eq}), the fidelity can be further expressed as 
 \begin{eqnarray}
 F_{EPR}
 &=&\frac{1-p}{P_{EPR} d_A^3 d_D} ~~\figbox{0.15}{FEPR.png}+\frac{p}{P_{EPR} d_A^3 d_D} ~~\figbox{0.15}{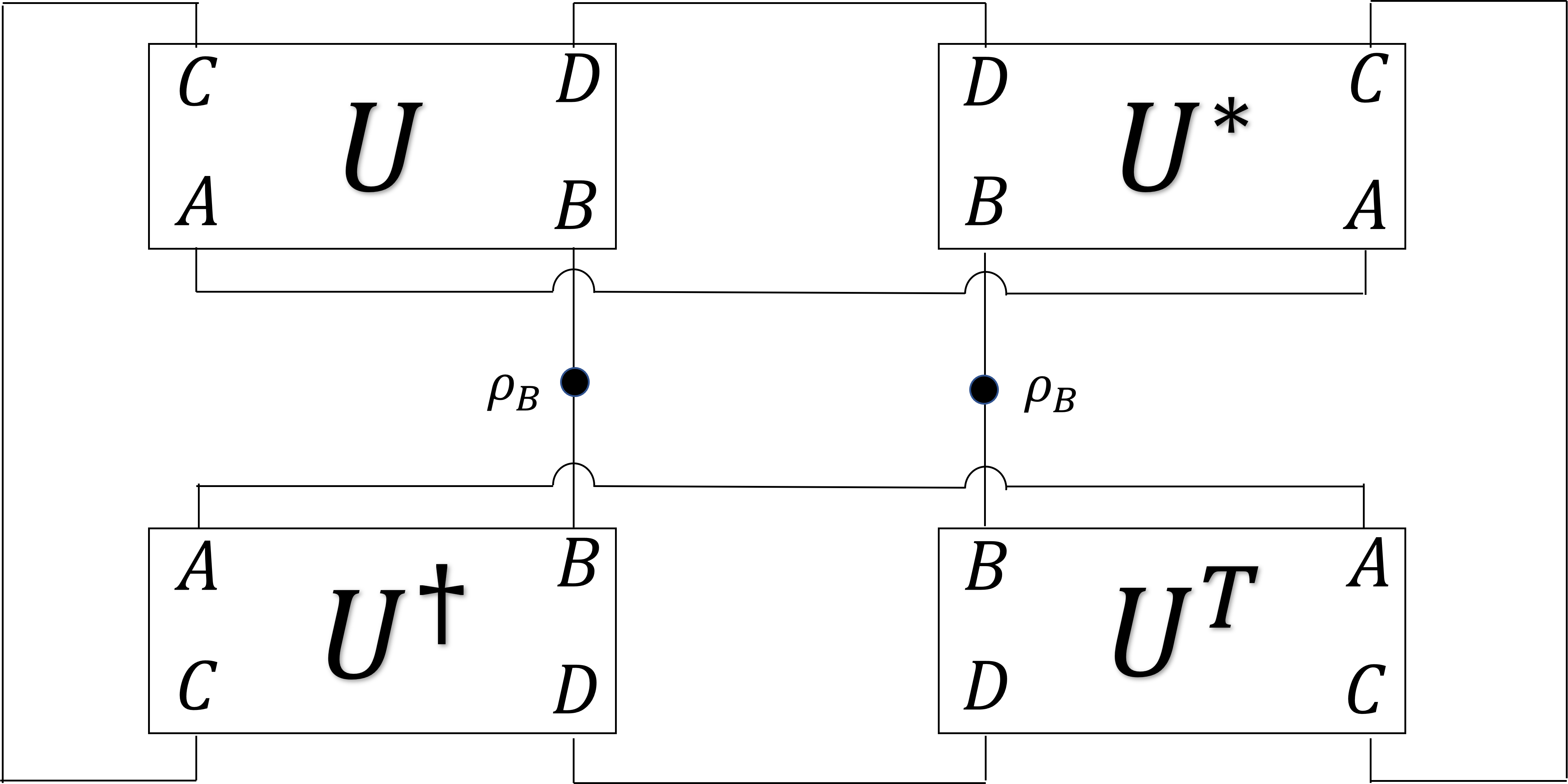}\nonumber\\
 &=&\frac{1-p}{P_{EPR} d_A^3 d_D} ~~~\figbox{0.2}{FEPR_cal.png}+\frac{p}{P_{EPR} d_A^3 d_D} ~~
 \figbox{0.2}{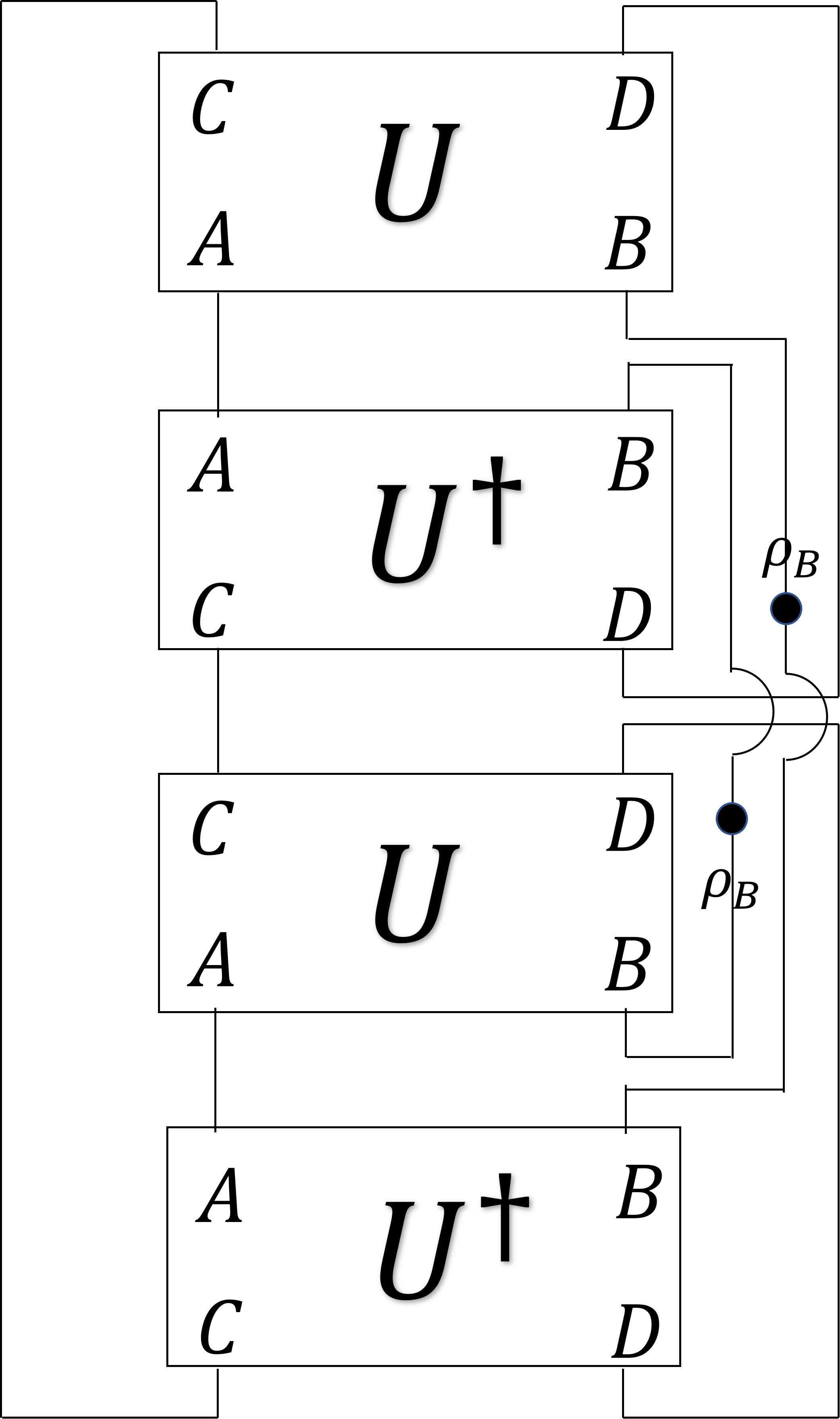}
 \end{eqnarray}
 The first graph has been calculated and the result is given by Eq.(\ref{F_EPR_ave}). The average value of the second graph can be explicitly calculated as  
\begin{eqnarray}
&&\int dU \left(\rho_{B}\right)_{b_1b_1'}
 \left(\rho_{B}\right)_{b_2b_2'}
  U_{c_1d_1a_1b_1}  U_{c_2d_2a_2b_2} 
 U^\ast_{c_2d_1a_1b_2'}  U^\ast_{c_1d_2a_2b_1'}\nonumber\\
&=&
\frac{d_A d_D}{(d^2-1)} 
 \left[d^2\langle\rho_B^2\rangle
 +d_C^2-d_C^2 \langle\rho_B^2\rangle-1
 \right]
\end{eqnarray}
 Therefore, the average value of the fidelity is given by 
 \begin{eqnarray}
  \bar{F}_{EPR}&=& \frac{1}{P_{EPR}d_A^2(d^2-1)} 
 \left[\left(d^2-d_C^2\right)\left(
 (1-p)d_B\langle\sqrt{\rho_B}\rangle^2+p\langle\rho_B^2\rangle
 \right)+ d_C^2-1\right]\nonumber\\
 &=&\frac{\left(d^2-d_C^2\right)\left(
 (1-p)d_B\langle\sqrt{\rho_B}\rangle^2+p\langle\rho_B^2\rangle
 \right)+ d_C^2-1}{\left(d^2-d_C^2\right)\left((1-p)d_B \langle\sqrt{\rho_B}\rangle^2
 +p \langle\rho_B^2\rangle\right)+d_A^2d_C^2-d_A^2}
 \nonumber\\
 &\approx& \frac{d_B^2 \left(
 (1-p)d_B\langle\sqrt{\rho_B}\rangle^2+p\langle\rho_B^2\rangle
 \right) +\frac{d_C^2}{d_A^2}}{d_B^2 \left(
 (1-p)d_B\langle\sqrt{\rho_B}\rangle^2+p\langle\rho_B^2\rangle
 \right) +d_C^2}
 \end{eqnarray}

It is shown that the decoding fidelity depends on the value of $(1-p)d_B\langle\sqrt{\rho_B}\rangle^2+p\langle\rho_B^2\rangle$. Obviously, the fidelity is less than unity. This indicates that the decoding strategy or the recovery algorithm of Hayden-Preskill protocol is harder to realize. If $d_B^2 ((1-p)d_B\langle\sqrt{\rho_B}\rangle^2+p\langle\rho_B^2\rangle)\gg d_C^2$, the fidelity approaches unity. The information can be extracted by using the decoding process successfully. If $d_B^2 ((1-p)d_B\langle\sqrt{\rho_B}\rangle^2+p\langle\rho_B^2 \rangle) \gg d_C^2$, the fidelity approaches $0$. In this limiting case, the information carried by the message system $A$ cannot be extracted by decoding the Hawking radiation.

\section{Conclusion and discussion}\label{conclusion}

In summary, we have studied the Hayden-Preskill thought experiment at finite temperature and obtained the decoupling condition that the information thrown into an old black hole can be extracted by decoding the Hawking radiation. We then consider the decoding Hayden-Preskill protocol at finite temperature assuming that the observer outside the black hole has the access to the full radiation and the unitary dynamics of the black hole. We also consider the cases when the Hawking radiation has noise and decoherence in the storage. The decoding probabilities and the corresponding fidelities are calculated. It is shown that for all the three cases we have considered, the decoding fidelities are less than unity in general. This result indicates that at finite temperature, the decoding strategy and the recovery algorithm is harder to realize than that at infinite temperature. However, there are some limiting cases that the fidelities approach to unity.

Recently, the decoding strategy of the Hawking radiation in the Hayden-Preskill thought experiment has been generalized to study the interior operator of the black hole \cite{Yoshida:2018ybz,Yoshida:2019kyp} and the implications in the black hole firewall paradox \cite{Yoshida:2019qqw}. One expects that our calculation may also have applications in this aspect. Another possible generalization of this work is to consider the case where the unitary dynamics is governed by the Clifford operator \cite{Yoshida:2021xyb} while the initial black hole and the early radiation are in the TFD state. These issues will investigated in the future works.


\begin{thebibliography}{}

\bibitem{Hawking:1975vcx}
S.~W.~Hawking,
``Particle Creation by Black Holes,''
Commun. Math. Phys. \textbf{43}, 199-220 (1975)
[erratum: Commun. Math. Phys. \textbf{46}, 206 (1976)].


\bibitem{Hawking:1976ra}
S.~W.~Hawking,
``Breakdown of Predictability in Gravitational Collapse,''
Phys. Rev. D \textbf{14}, 2460-2473 (1976).

\bibitem{Harlow:2014yka}
D.~Harlow,
``Jerusalem Lectures on Black Holes and Quantum Information,''
Rev. Mod. Phys. \textbf{88}, 015002 (2016)
[arXiv:1409.1231 [hep-th]].


\bibitem{Preskill:1992tc}
J.~Preskill,
``Do black holes destroy information?,''
[arXiv:hep-th/9209058 [hep-th]].

\bibitem{Susskind:1993if}
L.~Susskind, L.~Thorlacius and J.~Uglum,
``The Stretched horizon and black hole complementarity,''
Phys. Rev. D \textbf{48}, 3743-3761 (1993)
[arXiv:hep-th/9306069 [hep-th]].


\bibitem{Page:1993df}
D.~N.~Page,
``Average entropy of a subsystem,''
Phys. Rev. Lett. \textbf{71}, 1291-1294 (1993)
[arXiv:gr-qc/9305007 [gr-qc]].


\bibitem{Page:1993wv}
D.~N.~Page,
``Information in black hole radiation,''
Phys. Rev. Lett. \textbf{71}, 3743-3746 (1993)
[arXiv:hep-th/9306083 [hep-th]].

\bibitem{Hayden:2007cs}
P.~Hayden and J.~Preskill,
``Black holes as mirrors: Quantum information in random subsystems,''
JHEP \textbf{09}, 120 (2007)
[arXiv:0708.4025 [hep-th]].

\bibitem{Almheiri:2012rt}
A.~Almheiri, D.~Marolf, J.~Polchinski and J.~Sully,
``Black Holes: Complementarity or Firewalls?,''
JHEP \textbf{02}, 062 (2013)
[arXiv:1207.3123 [hep-th]].

\bibitem{Sekino:2008he}
Y.~Sekino and L.~Susskind,
``Fast Scramblers,''
JHEP \textbf{10}, 065 (2008)
[arXiv:0808.2096 [hep-th]].

\bibitem{Lashkari:2011yi}
N.~Lashkari, D.~Stanford, M.~Hastings, T.~Osborne and P.~Hayden,
``Towards the Fast Scrambling Conjecture,''
JHEP \textbf{04}, 022 (2013)
[arXiv:1111.6580 [hep-th]].

\bibitem{Shenker:2013pqa}
S.~H.~Shenker and D.~Stanford,
``Black holes and the butterfly effect,''
JHEP \textbf{03}, 067 (2014)
[arXiv:1306.0622 [hep-th]].

\bibitem{Maldacena:2015waa}
J.~Maldacena, S.~H.~Shenker and D.~Stanford,
``A bound on chaos,''
JHEP \textbf{08}, 106 (2016)
[arXiv:1503.01409 [hep-th]].

\bibitem{Gao:2016bin}
P.~Gao, D.~L.~Jafferis and A.~C.~Wall,
``Traversable Wormholes via a Double Trace Deformation,''
JHEP \textbf{12}, 151 (2017)
[arXiv:1608.05687 [hep-th]].

\bibitem{Maldacena:2017axo}
J.~Maldacena, D.~Stanford and Z.~Yang,
``Diving into traversable wormholes,''
Fortsch. Phys. \textbf{65}, no.5, 1700034 (2017)
[arXiv:1704.05333 [hep-th]].

\bibitem{Yoshida:2017non}
B.~Yoshida and A.~Kitaev,
``Efficient decoding for the Hayden-Preskill protocol,''
[arXiv:1710.03363 [hep-th]].

\bibitem{Yoshida:2018vly}
B.~Yoshida and N.~Y.~Yao,
``Disentangling Scrambling and Decoherence via Quantum Teleportation,''
Phys. Rev. X \textbf{9}, no.1, 011006 (2019)
[arXiv:1803.10772 [quant-ph]].


\bibitem{Yoshida:2018ybz}
B.~Yoshida,
``Soft mode and interior operator in the Hayden-Preskill thought experiment,''
Phys. Rev. D \textbf{100}, no.8, 086001 (2019)
[arXiv:1812.07353 [hep-th]].

\bibitem{Yoshida:2019qqw}
B.~Yoshida,
``Firewalls vs. Scrambling,''
JHEP \textbf{10}, 132 (2019)
[arXiv:1902.09763 [hep-th]].

\bibitem{Cheng:2019yib}
Y.~Cheng, C.~Liu, J.~Guo, Y.~Chen, P.~Zhang and H.~Zhai,
``Realizing the Hayden-Preskill protocol with coupled Dicke models,''
Phys. Rev. Res. \textbf{2}, no.4, 043024 (2020)
[arXiv:1909.12568 [cond-mat.quant-gas]].


\bibitem{Yoshida:2019kyp}
B.~Yoshida,
``Observer-dependent black hole interior from operator collision,''
Phys. Rev. D \textbf{103}, no.4, 046004 (2021)
[arXiv:1910.11346 [hep-th]].

\bibitem{Bao:2020zdo}
N.~Bao and Y.~Kikuchi,
``Hayden-Preskill decoding from noisy Hawking radiation,''
JHEP \textbf{02}, 017 (2021)
[arXiv:2009.13493 [quant-ph]].

\bibitem{Schuster:2021uvg}
T.~Schuster, B.~Kobrin, P.~Gao, I.~Cong, E.~T.~Khabiboulline, N.~M.~Linke, M.~D.~Lukin, C.~Monroe, B.~Yoshida and N.~Y.~Yao,
``Many-body quantum teleportation via operator spreading in the traversable wormhole protocol,''
[arXiv:2102.00010 [quant-ph]].

\bibitem{Hayata:2021kcp}
T.~Hayata, Y.~Hidaka and Y.~Kikuchi,
``Diagnosis of information scrambling from Hamiltonian evolution under decoherence,''
[arXiv:2103.05179 [quant-ph]].


\bibitem{Yoshida:2021xyb}
B.~Yoshida,
``Decoding algorithms for Clifford Hayden-Preskill problem,''
[arXiv:2106.15628 [quant-ph]].

\bibitem{Hosur:2015ylk}
P.~Hosur, X.~L.~Qi, D.~A.~Roberts and B.~Yoshida,
``Chaos in quantum channels,''
JHEP \textbf{02}, 004 (2016)
[arXiv:1511.04021 [hep-th]].

\bibitem{Wald:1975kc}
R.~M.~Wald,
``On Particle Creation by Black Holes,''
Commun. Math. Phys. \textbf{45}, 9-34 (1975).

\bibitem{Callan:1992rs}
C.~G.~Callan, Jr., S.~B.~Giddings, J.~A.~Harvey and A.~Strominger,
``Evanescent black holes,''
Phys. Rev. D \textbf{45}, no.4, R1005 (1992)
[arXiv:hep-th/9111056 [hep-th]].

\bibitem{Hayden:2008os}
P. Hayden, M. Horodecki, A. Winter, and J. Yard, “A decoupling approach to the quantum
capacity,” Open Syst. Inf. Dyn. 15 (2008) 7 [arXiv:quant-ph/0702005 [quant-ph]].


\bibitem{Russo:1992ax}
J.~G.~Russo, L.~Susskind and L.~Thorlacius,
``The Endpoint of Hawking radiation,''
Phys. Rev. D \textbf{46}, 3444-3449 (1992)
[arXiv:hep-th/9206070 [hep-th]].


\bibitem{Bose:1995pz}
S.~Bose, L.~Parker and Y.~Peleg,
``Semiinfinite throat as the end state geometry of two-dimensional black hole evaporation,''
Phys. Rev. D \textbf{52}, 3512-3517 (1995)
[arXiv:hep-th/9502098 [hep-th]].


\bibitem{Cruz:1997nj}
J.~Cruz, J.~Navarro-Salas, C.~F.~Talavera and M.~Navarro,
``Conformal and non-conformal symmetries in 2-D dilaton gravity,''
Phys. Lett. B \textbf{402}, 270-275 (1997)
[arXiv:hep-th/9606097 [hep-th]].

\bibitem{Cruz:1995zt}
J.~Cruz and J.~Navarro-Salas,
``Solvable models for radiating black holes and area preserving diffeomorphisms,''
Phys. Lett. B \textbf{375}, 47-53 (1996)
[arXiv:hep-th/9512187 [hep-th]].

\bibitem{Cruz:1996pg}
J.~Cruz and J.~Navarro-Salas,
``Black hole evaporation by thermal bath removal,''
Phys. Lett. B \textbf{387}, 51-56 (1996)
[arXiv:hep-th/9607155 [hep-th]].

\bibitem{Nielsen:2000}
M. A. Nielsen and I. L. Chuang, Quantum Computation and
Quantum Information (Cambridge University Press,
Cambridge, England, 2000).



\end{thebibliography}
\end{document}